\documentclass[12pt,preprint]{aastex}

\usepackage{color}

\newcommand{\SA}{semi-analytic}

\newcommand{\fValf}{f_\mathrm{alf}}

\newcommand{\Valf}{v_a}
\newcommand\Qlin{quasi-linear}

\newcommand{\Lres}{\lambda_\mathrm{res}}
\newcommand{\Kres}{k_\mathrm{res}}
\newcommand{\Blasi}{Blasi et al.}
\newcommand{\Berezhko}{Berezhko et al.}
\newcommand{\Utot}{U_\mathrm{tot}}
\newcommand{\MFA}{magnetic field amplification}
\newcommand{\BFA}{$B$-field amplification}

\newcommand{\Beff}{B_\mathrm{eff}}

\newcommand{\Dfeb}{d_\mathrm{FEB}}

\newcommand{\Qesc}{q_\mathrm{esc}}

\newcommand{\CR}{cosmic-ray}
\newcommand{\Diff}{D(x,p)}
\newcommand{\DiffPmax}{D(x,\Pmax)}

\newcommand{\Malf}{M_\mathrm{alf}} 
\newcommand{\MS}{M_\mathrm{s}} 
\newcommand{\alf}{Alfv\'en}

\newcommand{\Vwt}{v_\mathrm{wt}}

\newcommand{\MC}{Monte Carlo}

\newcommand{\Pp}{P_\mathrm{cr}}

\newcommand{\dPpxp}{\partial P_{p}(x,p)/\partial x}
\newcommand{\dPCRxp}{\partial P_\mathrm{cr}(x,p)/\partial x}

\newcommand{\dPpFxp}{\frac{\partial P_{p}(x,p)}{\partial x}}

\newcommand{\gamray}{$\gamma$-ray}

\newcommand{\NL}{nonlinear}
\newcommand{\muG}{$\mu$G}

\newcommand{\SC}{self-consistent}
\newcommand{\SCly}{self-consistently}

\newcommand{\pinj}{p_\mathrm{inj}}

\newcommand{\Pmax}{p_\mathrm{max}}
\newcommand{\TP}{test-particle}

\newcommand{\rg}{r_g}
\newcommand{\rgZ}{r_{g,0}}
\newcommand{\Pptot}{P_\mathrm{p,tot}}
\newcommand{\Fptot}{F_\mathrm{p,tot}}
\newcommand{\Pwtot}{P_\mathrm{w,tot}}
\newcommand{\Fwtot}{F_\mathrm{w,tot}}

\newcommand\Rtot{r_\mathrm{tot}}
\newcommand\Rsub{r_\mathrm{sub}}

\newcommand\Bcsm{B_\mathrm{CSM}}

\newcommand{\xx}[1]{\times 10^{#1}}

\newcommand{\myemail}{\texttt{don\_ellison@ncsu.edu}}
\newcommand{\AVemail}{\texttt{avladim@ncsu.edu}}
\newcommand{\rel}{relativistic}

\newcommand{\nonrel}{non-relativistic}
\newcommand{\transrel}{trans-relativistic}

\newcommand{\syn}{synchrotron}

\newcommand{\IC}{inverse-Compton}
\newcommand{\pion}{pion-decay}

\newcommand{\kmps}{km s$^{-1}$}
\newcommand{\pcc}{cm$^{-3}$}

\newcount\listnorom
\newcommand\listromanDE{\global\advance \listnorom by 1
{\lowercase\expandafter{(\romannumeral\listnorom)}\ }}
\newcommand\newlistroman{\listnorom=0}
\newcount\Inum
\newcount\IInum
\newcount\IIInum
\newcount\IVnum
\def\I{\global\multiply\IInum by 0 \global\multiply\IIInum by 0
            \global\multiply\IVnum by 0 \global\advance \Inum by 1
            {\the\Inum. }}
\def\II{\global\multiply\IIInum by 0\global\multiply\IVnum by 0
       \global\advance \IInum by 1 {\the\Inum.\the\IInum. }}
\def\III{\global\multiply\IVnum by 0\global\advance \IIInum by 1
            {\the\Inum.\the\IInum.\the\IIInum. }}
\def\IV{\global\advance \IVnum by 1
            {\the\IVnum. }}
\slugcomment{ApJ in press}

\shorttitle{Magnetic Field Amplification in NL DSA}
\shortauthors{Vladimirov, Ellison \& Bykov}

\begin{document}

\title{Nonlinear Diffusive Shock Acceleration with Magnetic Field
Amplification}

\author{Andrey Vladimirov and Donald C. Ellison} \affil{Physics
Department, North Carolina State University, Box 8202, Raleigh, NC
27695, U.S.A.} 
\author{Andrei Bykov} \affil{Department of Theoretical
Astrophysics, Ioffe Physical-Technical Institute, St. Petersburg RUSSIA}
\email{\AVemail, \myemail, byk@astro.ioffe.ru}

\begin{abstract}
We introduce a \MC\ model of \NL\ diffusive shock acceleration allowing
for the generation of large-amplitude magnetic turbulence, i.e., $\Delta
B \gg B_0$, where $B_0$ is the ambient magnetic field.  The model is the
first to include strong wave generation, efficient particle acceleration
to \rel\ energies in \nonrel\ shocks, and thermal particle injection in
an internally \SC\ manner.  We find that the upstream magnetic field
$B_0$ can be amplified by large factors and show that this amplification
depends strongly on the ambient \alf\ Mach number. We also show that in
the nonlinear model large increases in $B$ do not necessarily translate
into a large increase in the maximum particle momentum a particular
shock can produce, a consequence of high momentum particles diffusing in
the shock precursor where the large amplified field converges to the low
ambient value.
To deal with the field growth rate in the regime of strong fluctuations,
we extend to strong turbulence a parameterization that is consistent
with the resonant quasi-linear growth rate in the weak turbulence limit.
We believe our parameterization spans the maximum and minimum range of
the fluctuation growth and, within these limits, we show that the \NL\
shock structure, acceleration efficiency, and thermal particle injection
rates depend strongly on the yet to be determined details of wave growth
in strongly turbulent fields.  The most direct application of our
results will be to estimate magnetic fields amplified by strong
cosmic-ray modified shocks in supernova remnants.
\end{abstract}


\section{INTRODUCTION}
Recent observations and modeling of several young supernova remnants
(SNRs) suggest the presence of magnetic fields at the forward shock
(i.e., the outer blast wave) well in excess of what is expected from
simple compression of the ambient circumstellar field, $\Bcsm$.  These
large fields are inferred from
(i) spectral curvature in radio emission \citep[e.g.,][]{RE92,BKP99},
(ii) broad-band fits of \syn\ emission between
radio and non-thermal X-rays \citep[e.g.,][]{BKV2003,VBK2005a} \cite[see
also][]{Cowsik80},
and (iii) sharp X-ray edges
 \citep[e.g.,][]{VL2003,BambaEtal2003,VBK2005a,EC2005}.
While these methods are all indirect, fields greater than 500~\muG\ are
inferred in Cas A and values of at least several 100 \muG\ are estimated
in Tycho, Kepler, SN1006, and G347.3-0.5.
If $\Bcsm \sim 3-10$~\muG, amplification factors of 100 or more may be
required to explain the fields immediately behind the forward shocks and
this is likely the result of a nonlinear amplification process
associated with the efficient acceleration of cosmic-ray ions via
diffusive shock acceleration (DSA).
The magnetic field strength is a critical parameter in DSA and also
strongly influences the \syn\ emission from shock accelerated electrons.
Since shocks are expected to accelerate particles in diverse
astrophysical environments and \syn\ emission is often an important
emission process (e.g., radio jets), quantifying the magnetic field
amplification has become an important problem in particle astrophysics
and has relevance beyond cosmic-ray production in SNRs.

If \CR\ (CR) production is as efficient as expected in theories of
nonlinear DSA \citep[e.g.,][]{BE87,JE91,MD2001}, the CR pressure gradient
in the shock precursor can do work on the incoming plasma and, in
principle, place a large amount of energy in magnetic turbulence. In
fact, if the DSA process is to work at all, magnetic turbulence must be
self-generated on all resonant scale lengths to provide the scattering
necessary to drive the particle distribution to isotropy. If the
turbulence remains weak, i.e., if $\Delta B/B \ll 1$, the
self-generation of turbulence by CR streaming in the shock precursor
can be adequately described using
quasi-linear theory \citep[e.g.,][]{Skilling75a,MV82}. For large
amplitude turbulence, however, this analytic description becomes
questionable and less rigorous approximations must be made.

In principle, a complete description of \NL\ DSA, including \MFA, is
possible with particle-in-cell (PIC) simulations. In reality, however,
these simulations are still far too computationally demanding to produce
realistic models of astrophysical sources such as supernova
remnants. Approximate methods describing \NL\ DSA must be used.
Here, we have developed a method which incorporates a phenomenological
model of particle diffusion and turbulence growth \citep[similar
to][]{BL2001,AB2006} in a fully \NL\ \MC\ model of DSA
\citep[e.g.,][]{JE91,EBJ96,BE99}.  While not a first-principles
description of the plasma physics, the computational efficiency of our
approach gives it a significant advantage over PIC simulations
and \SA\ techniques based on the diffusion approximation in that
we can determine the shock structure, the injection of thermal
particles, the acceleration of these particles to \rel\ energies, and
the magnetic turbulence, all \SCly\ in a fully \NL\
steady-state model.

An important advantage of the \MC\ simulation is that, in effect, it
solves a more basic set of equations governing the shock structure and
DSA than do techniques based on diffusion--convection equations
\citep[e.g.,][]{EE84}.  There is no assumption of isotropy for particle
distributions and this allows an internally \SC\ treatment of thermal
particle injection.  Injection is still phenomenological and depends on
the assumptions made for the particle pitch-angle scattering, but these
assumptions are applied equally to all particles. Since there is only
one set of scattering assumptions, the \MC\ technique eliminates a free
``injection parameter'' which is present in all models based on the
diffusion approximation to set the injection efficiency. As we show, the
strong feedback between injection, shock structure, and \MFA\ makes this
property of the \MC\ technique particularly important.

Our preliminary results show that magnetic field amplification results
in an increase in the maximum CR momentum, $\Pmax$, a given
shock system can produce compared to the case without
amplification. 
For the case we consider, acceleration truncated by a finite-size shock,
this increase is not, however, as large as the increase
in $B$ in the shocked region since fields on scales of the upstream
diffusion length of the highest energy particles strongly influence
$\Pmax$. 
Furthermore, the fact that $B$ ranges smoothly from the unshocked value far
upstream, $B_0$, to the amplified value, $B_2$, downstream from the
shock will have consequences for electrons since radiation losses will
be greatest in the downstream region where particles spend 
a large fraction of their time.\footnote{We
consistently use the subscript 0 for far upstream values, the subscript
2 for downstream values, and the subscript 1 for values immediately
upstream from the subshock. Thus, the overall compression ratio is
$\Rtot = u_0/u_2$ and the subshock compression ratio is
$\Rsub=u_1/u_2$, where $u$ is the bulk flow speed.} 
Thus, the maximum momentum protons and electrons obtain may be
determined by two very different field strengths. 

We also show that the amplification factor, $B_2/B_0$, increases with
\alf\ Mach number, that is, for a given shock speed and ambient density,
a small $B_0$ will be amplified more than a larger $B_0$. The ability to
dramatically increase very low ambient fields may make it possible for
reverse shocks in young SNRs to accelerate particles and produce \rel\
electrons and radio \syn\ emission \cite[see][for a discussion of DSA at
reverse shocks is SNRs]{EDB2005}. The injection and acceleration of
electrons in amplified fields will be considered in subsequent
work, here, we consider only proton acceleration.

In this paper, we demonstrate the feasibility of applying \MC\
techniques to DSA with field amplification but have made a number of
approximations regarding wave generation and
have neglected wave damping.  An important advantage of the \MC\ method
is that it can be generalized to include more realistic descriptions of
non-resonant wave growth, linear and \NL\ damping, and the calculation
of the momentum and space dependent diffusion coefficient from the
turbulent energy density, and work along these lines is in progress.
While a more accurate description of the plasma physics will influence
all aspects of the acceleration process, we expect that $\Pmax$ 
will be most strongly determined by the plasma physics details and thus
will remain a critical problem for understanding the origin of galactic
cosmic rays for some time.

\section{MODEL}
\subsection{Assumptions for Magnetic Turbulence Generation}
Consider a steady-state collisionless shock propagating along a uniform
component ${\bf B_0}$ of a stochastic magnetic field ${\bf B}$.  We only
consider parallel shocks where ${\bf B_0}$ is along $x$ and the shock
face is perpendicular to ${\bf B_0}$.
As the unshocked plasma approaches the shock and experiences the
pressure gradient in the cosmic-ray precursor, 
the energetic particles backstreaming
from the shock 
cause fluctuations of the field, $\Delta {\bf B}$, to grow.
In the linear regime, $\Delta {\bf B}$ is perpendicular to ${\bf B}_0$
and the local rate of growth of energy in waves is proportional to the
particle pressure gradient. The plasma motion associated with these
field fluctuations is initially \alf ic: transverse and incompressible,
and it will remain as such as long as $\Delta B \ll B_0$.

As the perturbations grow and reach $\Delta B \gtrsim B_0$, however,
it is likely
that waves with wave vectors ${\bf k}$ not aligned with ${\bf B}_0$ will
be generated, due to local CR pressure gradients along the total ${\bf
B}={\bf B}_0 + \Delta{\bf B}$.  With $\Delta B \gtrsim B_0$, it becomes
impossible to predict the average value of the transverse pressure
gradients and the resulting magnetic field structure without knowing the
relative phases of different wave
harmonics.
The problem is further complicated by the fact that this longitudinal,
compressible turbulence may produce a strong second-order Fermi particle
acceleration effect which, in turn, can damp the longitudinal
fluctuations \citep[see, for example,][]{SchlickeiserEtal1993}. 

These complications place a precise description of plasma turbulence
beyond current analytic capabilities and lead to 
our goal of obtaining a realistic approximation
that includes the essential \NL\ effects from efficient DSA 
but still allows
particle acceleration over a large dynamic range.
In order to accomplish this,
we consider two limiting cases. The first assumes there is no
longitudinal turbulence, 
in which case the wave growth rate is determined by the \alf\ speed in
the un-amplified field $B_0$.
This gives a lower bound to the growth rate. 
The upper limit
assumes that the turbulence is isotropic, in which case the growth rate
is determined by the \alf\ speed in the much larger
amplified field $\Beff$ (defined below).  
The real situation should lie
between these two cases and while we consider these limits, we do not
explicitly include second-order Fermi acceleration in our calculations.

To get an initial estimate of the shock structure, we use a \MC\
simulation of diffusive shock acceleration, assuming that the motion of
particles can be described as pitch-angle scattering in the Bohm limit.
The \MC\ simulation provides the distribution of fast particles at all
positions relative to the viscous subshock and allows for a calculation
of the particle pressure gradient, $\dPpxp$, which drives the
amplification process in the shock precursor.\footnote{It's important to
note that while we speak of ``thermal'' and ``superthermal'' particles,
and may refer to superthermal particles as cosmic rays, the \MC\
simulation makes no distinction between thermal and superthermal
particles.  The same assumptions are applied to particles regardless
of their energy.}
Then, we use $\dPpxp$ to calculate the total energy density in magnetic
turbulence, $\Utot(x)$, as a function of position across the shock.

This calculation uses a semi-analytical expression for
wavelength-dependent growth of magnetic turbulent energy density $U(x,k)$,
which is similar to the well-known equation for \alf\ wave growth
\citep[e.g.,
equation B.8 in][]{MV82}. The complete expression is presented
below as equations~(\ref{uminuskp}) and (\ref{upluskp}). 
The turbulence amplification term in them, which
represents the streaming instability is:
\begin{equation}
\label{waveEq}
\left[ \frac{d}{dt} U(x,k) \right]_\mathrm{stream} =
V_G \left[ \dPpFxp \frac{dp}{dk} \right]_
{p=\bar{p}(k)}
\ ,
\end{equation}
where 
$p$ is the particle momentum, 
$\Utot(x) = \int_0^\infty{U(x,k)} dk$,
$\bar{p}(k)$ is the momentum of particles resonant with wavevector $k$,
and much of the complicated plasma physics is contained in $V_G$, an
unknown growth-rate coefficient with dimensions of speed.  Once the wave
energy density is determined (as described below), it is used to
calculate an ``effective'' amplified magnetic field, $\Beff(x) \equiv
\sqrt{4\pi \Utot(x)}$, and this field is then used to calculate the
particle diffusion coefficient, $\Diff$, as a function of position and
momentum.

For the growth of \alf\ waves in \Qlin\ theory, $V_G=\Valf$, where
$\Valf = B_0/\sqrt{4\pi \rho(x)}$ is the \alf\ speed calculated with the
un-amplified field and $\rho(x)$ is the matter density  at position $x$.
As mentioned above, taking $V_G=\Valf$ ignores the transverse gradients
of pressure and only the component of the pressure gradient along the
initial magnetic field $B_0$ is accounted for. This choice of $V_G$
provides a lower limit on the amplification rate and was used in
\citet{AB2006}.  If, on the contrary, we define $V_G$ using the
amplified field, i.e., $V_G = \Beff(x)/\sqrt{4 \pi \rho(x)}$, it
reflects the situation where the growth rate is determined by the
maximum gradient of $P_p(x,p)$ along the fluctuating field lines. This
provides an upper limit on the wave growth rate and was used in
\citet{BL2001}.

A similar argument applies to the resonance condition. 
In order to obtain a solution, we assume a resonance relation between
$k$ and $p$ even though there is unlikely to be such a simple relation
in strong turbulence.  Again, we can define two limiting cases where the
resonant wavelength for particles of momentum $p$ lies between $\Lres(p)
= 2\pi \rg(B_0,p)$ and $\Lres(p) = 2\pi \rg(\Beff,p)$, where $\rg =
pc/(eB)$ is the gyroradius.
The real situation should lie between the two extremes for $V_G$ and
$\Lres$.
For this preliminary work, we set $\Lres(p) = 2\pi \rg(B_0,p)$, but vary
$V_G$ between the two limits, i.e., we introduce a parameter, $0 \le
\fValf \le 1$, such that
\begin{equation}
\label{VG}
V_G = \Valf \left \{ 1 + \left [\frac{\Beff(x)}{B_0} -1 \right ] 
\fValf \right \}
\ ,
\end{equation}
and $V_G$ varies linearly between $\Valf$ (for $\fValf=0$) and
$\Beff/\sqrt{4 \pi \rho(x)}$ (for $\fValf=1$).

Finally, we assume a Bohm model for diffusion. The mean free path of a
particle with momentum $p$ at position $x$ is taken to be equal to the
gyroradius of this particle in the amplified field, i.e., $\lambda(x,p)
= r_g(x,p) = pc/[q \Beff(x)]$, and the diffusion coefficient is then
$\Diff = \lambda v/3$, where $v$ is the particle speed.
Closure of the model is provided by using the newly calculated diffusion
coefficient in the \MC\ simulation, calculating a new shock structure
and pressure gradient profile $\dPpxp$, and iterating until mass,
momentum, and energy fluxes are conserved throughout the shock.

\subsection{Monte Carlo Simulation}
\label{MCsection}
The basic \MC\ (MC) simulation used here is described in detail in a number
of papers including \citet{JE91} and \citet{EJR90,EJB99}.  The model
calculates the shock structure and \NL\ particle spectrum \SCly,
including thermal particle injection. As mentioned above, injection can
be treated \SCly\ in this model because the \MC\ technique does not
require the diffusion approximation and no distinction between thermal
and superthermal particles is made.
Until now, however, in parallel shocks we have assumed a spatially
independent form for the diffusion coefficient, i.e., that the diffusion
coefficient, $D(p)$, is proportional to gyroradius: $D(p) = \lambda
v/3$, where $\lambda = \eta \rgZ$ is the particle mean free path, $\rgZ
= pc/(qB_0)$ is the gyroradius, and $\eta \ge 1$ is an arbitrary
parameter determining the ``strength'' of scattering.\footnote{Parallel
shocks are those where the shock normal is parallel to the magnetic
field direction and oblique shocks are ones where the field makes some
angle to the normal. The MC simulation has been generalized for plane,
oblique shocks, in which case $B$ and $D$ vary with $x$ as the strength
and angle the magnetic field makes with the shock normal vary
\citep[e.g.,][]{EBJ96,EJB99,ED2004}. }
In fact, we assume $\lambda$ and then individual particles pitch-angle
scatter with some maximum scattering angle which is set to give the
assumed mean free path \citep[see][for full details of the pitch-angle
scattering process]{EJR90,ED2004}.

The \MC\ simulation obtains the shock structure or bulk flow speed
$u(x)$ by iteration, ensuring that mass, momentum, and energy fluxes are
conserved across the shock.  When DSA is efficient, the shock is
modified by the back pressure of accelerated particles and the flow
speed, $u(x)$, becomes a ``smooth'' function of $x$, i.e., an upstream
precursor with $P_p \sim \rho u^2$ forms.
In addition to shock smoothing, the overall compression ratio of the
shock, $\Rtot$, will increase when the acceleration is efficient
\citep[e.g.,][]{Eichler84,JE91} and $\Rtot$ is obtained by iteration as
well.  Below, we show how our code has been generalized to include the
modification of the diffusion coefficient by the buildup of magnetic
turbulence.

\subsection{Magnetic Field Amplification}
\label{Sect_Bamp}
To calculate the effect of the pressure gradient of particles on
magnetic field fluctuations, we start with an equation similar to
equation~(B.8) in \citet{MV82}:
\begin{equation}
\label{McKVeqn}
\frac{\partial}{\partial t} U_w +
\frac{\partial}{\partial x} F_w = u
\frac{\partial}{\partial x} P_w
- v_{a,x}\frac{\partial \Pp }{\partial x} - \bar{L}
\ .
\end{equation}
Here $U_w=(\Delta B)^2/4\pi$ is the energy density of the waves
(assuming that the kinetic energy density of shear wave motion equals
the magnetic energy density), $F_w=( 3 u/2 - v_a ) U_w$ is the wave
energy flux (the $3/2$ represents the sum of the Poynting flux and the
flux associated with the transverse motion of plasma in \alf\ waves),
$P_w=U_w/2$ is the magnetic pressure of waves acting on the plasma flow,
and $\bar{L}$ represents wave energy losses (or gains) due to processes
other than compression and amplification by the considered instability,
and $\partial \Pp /\partial x$ is the pressure gradient of cosmic rays
exciting the \alf\ waves. This equation describes the growth of the
energy in \alf\ waves through their instability in the presence of CR
streaming, and it assumes that all the waves are moving upstream with
respect to the plasma at speed $v_a$, so $v_{a,x}=-v_a$.

Now, following \citet{BL2001}, we separate the turbulence into
downstream- and upstream-moving structures, and define the energy
density of these structures per wavenumber interval as $U_+(x,k)$ and
$U_-(x,k)$, respectively, so that the total energy density of turbulence
is
\begin{equation}
\label{utotal}
\Utot = \int_0^{\infty} [U_-(x,k)+U_+(x,k)]dk
\ .
\end{equation}
We also define the partial pressure of particles with momentum $p$ per
unit momentum interval as $P_{p}(x,p)$ so that the total pressure in
particles, including thermal ones, is
\begin{equation}
\label{ptotal}
\Pptot = \int_0^{\infty} P_{p}(x,p) dp
\ .
\end{equation}

To derive equations for $U_{\pm}(x,k)$, we apply a steady-state version
of (\ref{McKVeqn}) to waves with wavenumber $k$ in the interval $\Delta
k$.  For the energy density of these waves in the first order of $\Delta
k$ we substitute
\begin{equation}
U_w=\left[ U_+(x,k)+U_-(x,k) \right] \Delta k
\ .
\end{equation}
For the energy flux, we write
\begin{equation}
F_w = \left[ \left( \frac32 u(x) - V_G \right)U_-(x,k)+
             \left( \frac32 u(x) + V_G \right)U_+(x,k) \right] \Delta k
\ ,
\end{equation}
and for the pressure of particles interacting with these waves,
\begin{equation}
\Pp = P_{p}(x,p) \Delta p
\ .
\end{equation}
Substituting these expressions into (\ref{McKVeqn}), ignoring the
energy loss (gain) term $\bar{L}$, then dividing both
sides by $\Delta k$ and taking the limit $\Delta k \to 0$, we get
\begin{displaymath}
\frac{\partial}{\partial x} \left[ 
\left( \frac32 u(x) - V_G \right) U_-(x,k) +
\left( \frac32 u(x) + V_G \right) U_+(x,k) \right] = 
\qquad \qquad \qquad \qquad
\end{displaymath}
\begin{equation}
\label{amplinnew}
 \qquad \qquad \qquad \qquad 
u(x) \frac{\partial}{\partial x} \left(
\frac12 U_-(x,k) + \frac12 U_+(x,k) \right) -\Vwt\frac{\partial
\Pp(x,p)}{\partial x} \left| \frac{dp}{dk} \right| \ ,
\end{equation} 
where we have replaced $v_{a,x}$ with a weighted wave speed 
\begin{equation}
\Vwt(x,k) \equiv  V_G \frac{U_+(x,k) - U_-(x,k)}{U_+(x,k) + U_-(x,k)}
\ ,
\end{equation}
in the driving term.
Using $\Vwt$ is a logical extension of the definition of an ``average''
wave speed by \citet{BL2001} applied to a narrow wavenumber range
$\Delta k$.  This, again, is justified as long as $\Delta B \ll B_0$,
but becomes less clear for strong turbulence.  In these equations, $V_G$
is defined as in equation~(\ref{VG}).

The coefficient $|dp/dk|$ that the driving term has acquired is
necessary to relate the interval of wavenumbers $\Delta k$ of amplified
waves to the interval of momenta of particles interacting with these
waves $\Delta p$.  As mentioned above, the relationship between $p$ and
$k$ (i.e., the resonant condition) is assumed to be $k=1/ \rgZ$, where
$\rgZ = cp/(e B_0)$, and $B_0$ is the far upstream magnetic field. 
One may argue that the field that the particles `feel' as uniform in the
non-linear regime $\Delta B \gg B_0$ is the field carried by all
long-wavelength (relative to the particle gyroradius) harmonics, rather
than $B_0$.  While this may be true in general, our present model is
insensitive to the choice of the resonance condition due to the
simplified ``Bohm'' form of the particle mean free path we assume.

Equation (\ref{amplinnew}) expresses energy conservation and accounts
for the amplification of turbulence by the streaming instability with the
growth rate determined from kinetic theory. 
Furthermore, we assume, as did \citet{BL2001}, that interactions between
the forward and backward moving waves drive them to isotropy on a time
scale 
$\sim \rgZ/V_G$. In order to account for this interaction, we write
equation~(\ref{amplinnew}) as the sum of the following two equations for
$U_{\pm}(x,k)$:
\begin{displaymath}
[u(x) - V_G]\frac{\partial}{\partial x}U_- + U_-\frac{d}{dx} \left(
\frac32 u(x) - V_G \right) = 
\qquad\qquad\qquad\qquad\qquad
\end{displaymath}
\begin{equation}
\label{uminuskp}
\qquad\qquad\qquad\qquad\qquad \frac{U_-}{U_+ + U_-} V_G \frac{\partial
    \Pp(x,p)}{\partial x}\left|\frac{dp}{dk}\right|- \frac{V_G}{\rgZ}
    \left( U_- - U_+ \right)
\ ;
\end{equation}
\begin{displaymath}
[u(x) + V_G]\frac{\partial}{\partial x}U_+ + U_+\frac{d}{dx} \left(
\frac32 u(x) + V_G \right) = 
\qquad\qquad\qquad\qquad\qquad
\end{displaymath}
\begin{equation}
\label{upluskp}
\qquad\qquad\qquad\qquad\qquad -\frac{U_+}{U_+ + U_-} V_G
\frac{\partial \Pp(x,p)}{\partial x}\left|\frac{dp}{dk}\right|+
\frac{V_G}{\rgZ} \left( U_- - U_+ \right)
\ ,
\end{equation}
which are solved iteratively in the MC simulation.
We note that equations~(\ref{uminuskp}) and (\ref{upluskp}) are
consistent with equation~(\ref{amplinnew}) with or without the relaxation terms
on the right-hand sides of both equations, but these terms may become
important in cases with small shock velocities.

Equations (\ref{uminuskp}) and (\ref{upluskp}) are generalizations of
those introduced by \citet{BL2001}. The generalization has two
essential improvements. First, it accounts for the spatial dependence of
flow speed $u(x)$ due to \NL\ effects of efficient DSA and,
consequently, treats `compression' of the amplified field adequately. To
illustrate this effect, consider (\ref{uminuskp}) and (\ref{upluskp}),
neglecting the cosmic-ray pressure gradient term and $v_a$, i.e., taking
$V_G \ll u$. Adding the
two equations then results in
\begin{equation}
u(x)\frac{\partial}{\partial x}[U_-+U_+]+\frac32 [U_-+U_+] 
\frac{d}{dx}u(x)=0
\ ,
\end{equation}
which can be easily integrated to give $[U_-+U_+] \propto u^{-3/2}$, or
$\Beff \propto u^{-3/4}$. Consequently, in a shock with a total
compression ratio $\Rtot=10$, for example, the stochastic magnetic field
gets a boost in amplification by a factor of about $\Rtot^{-3/4} \simeq 6$ solely through
the compression of the plasma.
This compressional effect is especially important at the subshock and
 the change in magnetic turbulence energy density across the subshock
 will influence the subshock compression ratio $\Rsub$. This in turn
 will have a strong effect on the injection efficiency in the Monte
 Carlo model.

Second, we generalized the equations to describe the whole spectrum of
turbulence $U_{\pm}(x,k)$ rather than a single waveband with $\Delta
k=k$. We solve this system with a finite-difference method, integrating
from far upstream ($x \to -\infty$) to $x$. The quantities $u(x)$ and
$P_{p}(x,p)$ are obtained from the \MC\ simulation, as described in
Section \ref{MCsection}.
For simplicity in this initial presentation of our model, and because
the shocks we are mainly concerned with have high \alf\ Mach numbers
($v_a \ll u$), we have neglected $V_G$ compared to $u$ in the first and
second terms in equations~(\ref{uminuskp}) and (\ref{upluskp}) in the
numerical results we present here.\footnote{In all of the examples in
  this paper, $V_G < 0.2 u(x)$ in the shock precursor and $V_G < 0.5
  u_2$ in the post-shock region.}
We do, however, account for the wave or scattering center speed relative
to the bulk flow speed in determining the energy change a particle
receives as it scatters in the converging flow. In each interaction, we
replace $u(x) \rightarrow u(x) + \Vwt$, and since $\Vwt$ is generally
negative in the upstream region (i.e., $U_->U_+$), a finite $\Vwt$
dampens particle acceleration. Downstream from the shock we take
$\Vwt=0$.

The initial condition for our equations is the far upstream magnetic
turbulence spectrum $U(x \to \infty, k)$. We take
\begin{equation}
\label{boundary}
U_-(x\to -\infty,k)=U_+(x\to -\infty,k)=
\left\{
\begin{array}{cl}
A ( k/k_c) ^{-\alpha}, & k_c<k<k_m \\
0, & k<k_c \quad {\rm or} \quad k > k_m 
\ ,
\end{array}\right.
\end{equation}
where the limits $k_m$ and $k_c$ are chosen to encompass the range
between the inverse gyroradii of thermal particles and the most
energetic particles in the shock, respectively.  For concreteness we
take $\alpha=1$, but none of our results depend in any substantial way
on $\alpha$, $k_m$, or $k_c$.
Using our definition of 
the effective, amplified magnetic field,
\begin{equation}
\label{beff}
\Beff^2(x)=4\pi \int_{0}^{\infty}[U_-(x,k)+U_+(x,k)]dk
\ ,
\end{equation}
the normalization constant $A$ in equation~(\ref{boundary}) is
determined by requiring that
\begin{equation}
4\pi\int_{0}^{\infty}[U_-(-\infty,k)+U_+(-\infty,k)]dk=B_0^2
\ .
\end{equation}

\subsection{Particle Scattering}
\label{Sect_scat}
Once the amplified magnetic field is determined from
equation~(\ref{beff}), the scattering mean free path is set equal to the
gyroradius of the particle in this field, i.e.,
\begin{equation}
\label{mfp}
\lambda(x,p) = \frac{cp}{e \Beff(x)}
\ .
\end{equation}
Equation~(\ref{mfp}) is essentially the Bohm limit and, as such, is 
a crude approximation.
In a strong collisionless shock, modified by efficient DSA, a
significant fraction of the energy is contained in high-momentum
particles.
These particles have long mean free paths and will resonantly produce
turbulence where long-wavelength harmonics contain  most of the wave
energy.
Consequently, to low-momentum particles, the strong long-wavelength
turbulence appears approximately as a uniform field and
equation~(\ref{mfp}) is justified.
For the highest energy particles, however, equation~(\ref{mfp}) will
overestimate the scattering strength since, for these particles, most of
the harmonics of the magnetic field appear as short-scale fluctuations
which are not very efficient at changing the particle's momentum.
In this case, there is no reason to assume that the scattering is
resonant, i.e., there is no simple resonance relation between $k$ and $p$.
This is a critical point since the form for $\lambda(x,p)$ at high $p$
determines the maximum momentum that can be produced in a given shock
system, and this is one of the important unsolved problems for DSA.

Clearly, more physically realistic models for both the diffusion
coefficient and the resonance condition are required for future work. An
approach for determining the mean free path of particles in strongly
turbulent fields is described in \citet{BT92}, where non-resonant
scattering and diffusive transport of particles in large-scale
fluctuations are taken into account. The model presented here, where the
calculation of the power spectrum of turbulence $U_{\pm}(x,k)$ is
coupled to the \NL\ shock structure, will be generalized to include more
physically realistic wave-particle interactions in future work.

\subsection{Momentum and Energy Conservation}
The total energy density of the MHD turbulence is defined by
equation~(\ref{utotal}), and it is assumed to be equally shared between
the stochastic magnetic field and the stochastic incompressible
transverse motion of the gas. The momentum flux of the turbulence is
then the magnetic pressure, i.e.,
\begin{equation}
\Pwtot(x)=\frac12 \int_0^{\infty} [U_-(x,k)+U_+(x,k)] dk = 
\frac12 \Utot(x)
\ .
\end{equation}
The total energy flux in turbulence is
\begin{equation}
\Fwtot(x)=\int_0^{\infty} \left[ 
  \left(\frac32 u - V_G \right) U_-(x,k) +
  \left(\frac32 u + V_G \right) U_+(x,k) \right] dk 
\ ,
\end{equation}
or
\begin{equation}
\Fwtot(x) \approx 
\frac32 u(x) \Utot(x)
\ ,
\end{equation}
in the limit $V_G(x) \ll u(x)$.

In order to determine the bulk velocity profile, $u(x)$, consistent with the
back-reaction of accelerated particles and turbulence on the flow, the
Monte Carlo simulation solves the following equations expressing the 
conservation of mass and momentum fluxes, i.e.,
\begin{eqnarray}
\label{massflux}
\rho(x)  u(x) &=& \rho_0 u_0 \, , \\
\label{pflux}
\rho(x) u(x)^2 + \Pptot(x) + \Pwtot(x) &=& \rho_0 u_0^2 + P_{p0} + P_{w0}
\equiv P_0
\ ,
\end{eqnarray}
where $\Pptot(x)$ is the pressure produced by all particles, thermal and
superthermal and $P_0$ is the far upstream momentum flux.  The particle
pressure is related to $P_{p}(x,p)$ by equation~(\ref{ptotal}), and is
calculated in the simulation directly from the trajectories of
individual particles.
The energy flux conservation relation is
\begin{equation}
\label{engyflux}
\frac{\rho(x) u(x)^3}{2} + \Fptot(x) + \Fwtot(x) + \Qesc =
      \frac{\rho_0 u_0^3}{2}+F_{p0} + F_{w0} \equiv
F_0
\ ,
\end{equation}
where $\Fptot(x)$ is the energy flux in all particles, $\Qesc$ is the
energy flux lost by particles leaving the system at the upstream free
escape boundary (FEB), and $F_0$ is the far upstream
energy flux.
In parallel shocks, $u(x)$ can be determined from
equations~(\ref{massflux}) and (\ref{pflux}) alone.  The relation
between $\Rtot$ and $\Qesc$ \citep[i.e., eq.~10 in][]{EMP90} allows
equation~(\ref{engyflux}) to be used to check the consistency of the
simulation results, as we show with the examples below.

\section{RESULTS}
In all of the following examples we set the shock speed
$u_0=5000$\,\kmps, the unshocked proton number density $n_{p0}=1$\,\pcc,
and the unshocked proton temperature $T_0=10^6$\,K. For simplicity, the
electron temperature is set to zero and the electron contribution to
the jump conditions is ignored. With these parameters, the sonic Mach
number $\MS \simeq 43$ and the \alf\ Mach number $\Malf \simeq 2300
(1\mu \mathrm{G}/B_0)$.

\subsection{With and Without Magnetic Field Amplification}
In order to obtain a solution which conserves momentum and energy
fluxes, both the shock structure, i.e., $u(x)$, and the overall
compression ratio, $\Rtot$, must be obtained \SCly.  
Figure~\ref{BL_noBL} shows results for four shocks where $u(x)$ and
$\Rtot$ have been determined with the iterative method described
above. In all examples in this section, $\fValf=0$.
First, we compare the results shown with heavy-weight solid curves to
those shown with heavy-weight dotted curves.  The heavy solid curves
were determined with $B$-field amplification while the dotted curves
were determined with a constant $\Beff(x)=B_0$. All other input
parameters were the same for these two models, i.e., $B_0=30$\,\muG,
$\fValf=0$ (i.e., minimum wave amplification), and an upstream free
escape boundary at
$\Dfeb=-10^4\,\rg(u_0)$, where $\rg(u_0) \equiv m_p u_0 c/(eB_0)$.
The most striking aspect of this comparison is the increase in
$\Beff(x)$ when field amplification is included (bottom panels). The
magnetic field goes from $\Beff(x\to -\infty) = 30$\,\muG, to $\Beff >
1000$\,\muG\ for $x > 0$, and this factor of $>30$ increase in $B$ will
influence the shock structure and the particle distributions in
important ways.
Note that there is about a factor of $\sim 2$ jump in $B$ at the
subshock at $x = 0$. As shown in Section~\ref{Sect_Bamp}, the jump in $B$ at the
subshock from compression (ignoring the contribution from the particle
pressure gradient) is $B_2/B_1 \sim (u_1/u_2)^{3/4} = \Rsub^{3/4}$,
where $u_1$ is the flow speed immediately upstream from the subshock and
the subshock compression ratio is defined as $\Rsub=u_1/u_2$. For the
heavy-weight solid curve in Figure~\ref{BL_noBL}, $\Rsub \simeq 2.7$ and
$B_2/B_1 \sim 2$, as observed.
 
The solution without $B$-field amplification (dotted curves) has a
considerably larger $\Rtot$ than the one with amplification, i.e., for
no $B$-field amplification, $\Rtot \simeq 22$, and with $B$-field
amplification (heavy solid curves), $\Rtot \simeq 11$.\footnote{See
\citet{BE99} for a discussion of how very large $\Rtot$'s can result in
high Mach number shocks if only adiabatic heating is included in the
precursor. The uncertainty on the compression ratios for the examples in
this paper is typically $\pm 10\%$.}
This
 difference in overall compression results because the wave pressure
 $\Pwtot$ in equation~(\ref{pflux}) is much larger in the field
 amplified case making the plasma less compressible.  

Two effects cause $\Rtot$ to increase above the \TP\ limit of 4 for
 strong shocks.  The first is the production of \rel\ particles which
 produce less pressure for a given energy density than \nonrel\
 particles making the plasma more compressible. 
The second, and most important, is the escape of energetic
 particles at the FEB.  As indicated in the energy flux panels of
 Figure~\ref{BL_noBL}, the energy flux drops abruptly at $x \sim \Dfeb$ as
 energetic particles diffuse past the FEB and escape the system.
The energy lost from the escaping particles is analogous to radiation
 losses in radiative shocks and results in an increase in compression
 ratio.  For high Mach number parallel shocks, $1/\Rtot \sim [5 - (9 +
 16 \Qesc/F_0)^{1/2}]/8$ \citep[][]{EMP90}, so for the dotted curves
 $\Rtot = 22$ and $\Qesc \simeq 0.8 F_0$, and for the heavy solid curves
 $\Rtot = 11$ and $\Qesc \simeq 0.5 F_0$, consistent with the energy
 fluxes shown in Figure~\ref{BL_noBL}.

The momentum flux is also conserved, but the escaping momentum flux is a
much smaller fraction of the upstream value than for energy
\citep[][]{Ellison85}. Any departures from the far upstream momentum
flux seen in Figure~\ref{BL_noBL} are less than the statistical
uncertainties in the simulation.
As mentioned above, we only include adiabatic heating in the upstream
precursor. If wave damping and heating were included, it is expected
that $\Rtot$ would be reduced from what we see with wave amplification
alone.

For a given shock, an increase in $\Rtot$ must be accompanied by a
 decrease in the subshock compression ratio, $\Rsub$ \citep[see, for
 example,][for a discussion of this effect]{BE99}.
A large $\Rtot$
means that high energy particles with long diffusion lengths get accelerated
very efficiently and, therefore, the fraction of particles injected must
decrease accordingly to conserve energy. The shock structure adjusts so
weakened injection (i.e., a small $\Rsub$) just balances the more efficient
acceleration produced by a large $\Rtot$. Since $\Rsub$ largely
determines the plasma heating, the more efficiently a shock accelerates
particles causing $\Rtot$ to increase, the less efficiently the plasma
is heated.
For the cases shown in Figure~\ref{BL_noBL}, $\Rsub \simeq 3.8$ for the
amplified field case (heavy solid curves) and $\Rsub\simeq 2.4$ for
the case with no $B$-field amplification (heavy dotted curves).

In Figure~\ref{BL_noBL_fp} we show the phase space distributions, $f(p)$,
for the shocks shown in Figure~\ref{BL_noBL}.  For the two cases with the
same parameters except field amplification, we note that the amplified
field case (heavy solid curve) obtains a higher $\Pmax$ and has a higher
shocked temperature (indicated by the shift of the ``thermal'' peak and
caused by the larger $\Rsub$) than the case with no field amplification
(heavy dotted curves).
It is significant that the increase in $\Pmax$ is modest even though $B$
increases by more than a factor or 30 with field amplification.
We emphasize that $\Pmax$ as such is not a parameter in this model;
$\Pmax$ is determined \SCly\ once the size of the shock system, i.e.,
$\Dfeb$, and the other environmental parameters are set.

In order to show the effect of changing $\Dfeb$, we include in
Figs.~\ref{BL_noBL} and \ref{BL_noBL_fp} field amplification shocks with
the same parameters except that $\Dfeb$ is changed to $-1000\,\rg(u_0)$
(dashed curves) and $-10^5\,\rg(u_0)$ (light-weight solid curves).
From Figure~\ref{BL_noBL_fp}, it's clear that $\Pmax$ scales approximately
as $\Dfeb$ and that the concave nature of $f(p)$ is more pronounced for
larger $\Pmax$.  The field amplification also increases with $\Pmax$,
but the increase between the $\Dfeb=-1000\,\rg(u_0)$ and $\Dfeb =
-10^5\,\rg(u_0)$ cases is less than a factor of two 
(bottom panels of Figure~\ref{BL_noBL}).

In Figure~\ref{U_D_fp} we show the energy density in magnetic turbulence,
$U_+(x,k) + U_-(x,k)$, the diffusion coefficient, $D(x,p)$, 
and particle distributions as functions of $k$ and $p$ at three different
positions in the shock. All of these plots
are for the example shown with dashed curves in Figs.~\ref{BL_noBL}
and \ref{BL_noBL_fp}.  The bottom panel shows how $f(x,p)$ varies in the
precursor where particles must diffuse upstream against the incoming
plasma.
The upstream diffusion length, $[\Diff/u(x)]_\mathrm{ave}$, is some
weighted average which is determined directly in the \MC\ simulation.
The diffusion coefficients, shown in the middle panel, are determined
from $\Beff$ (equation~\ref{beff}), where the $U(x,k)$'s (top panel) are
determined by solving equations~(\ref{uminuskp}) and (\ref{upluskp})
\SCly\ with the shock structure using the particle pressure gradients
determined from $f(x,p)$.  The decrease in $D$ by $\sim 40$
between far upstream and downstream from the shock corresponds to the
increase in $\Beff$ shown with the dashed curve in the bottom panel of
Figure~\ref{BL_noBL}.
The top panel clearly shows the spread in $k$ where resonant
interactions produce wave growth at the different $x$-positions,
including the portion from thermal particles at high $k$.

The efficiency of the shock acceleration process is shown in
Figure~\ref{inj_eff}. The curves on the left give the number density of
particles with momentum greater than $p$, i.e.,  $N(>p)$, and the curves on
the right give the energy density in particles with momentum greater
than $p$, i.e.,  $E(>p)$.  
Both sets of curves 
are determined solely from the downstream particle distributions
(calculated in the shock reference frame) shown in Figs.~\ref{BL_noBL}
and \ref{BL_noBL_fp} with heavy solid and dotted curves. Thus they
ignore escaping particles and the energy in magnetic turbulence.  
Nevertheless, these curves indicate that the shocks are extremely
efficient accelerators with $>50\%$ of the energy density in $f(p)$
placed in \rel\ particles (i.e., $p \ge m_pc$). The actual energy
efficiencies are considerably higher since the escaping particles carry
away a larger fraction of the total energy than is placed in magnetic
turbulence. 
With $\Qesc$ included, well over 50\% of the
total shock energy is placed in \rel\ particles. Despite this high
energy efficiency, the fraction of total particles that become \rel\ is
small, i.e., $N(>p=m_pc) \sim 10^{-5}$ in both cases.

The effect of magnetic field amplification on the number of particles
injected is evident in the left-hand curves. The larger $\Rsub$ (solid
curve) results in more downstream particles being injected into the
Fermi mechanism with amplification than without. 
While it is hard to see from Figure~\ref{inj_eff}, when the escaping
energy flux is included, the shock with \BFA\ puts a considerably
smaller fraction of energy in \rel\ particles than the shock without
amplification.  Again, injection depends in a \NL\ fashion on the shock
parameters and the subshock strength will adjust to ensure that just the
right amount of injection occurs so that momentum and energy are
conserved.

\subsection{Alfv\'en Mach Number Dependence}
In Figure~\ref{vary_Bz} we show three examples where $B_0$ has been varied
from 0.3 to 3 to 30~\muG; all other input parameters are kept constant
including the physical distance to the FEB and $\fValf=0$. For these
examples, $|\Dfeb| = 1.7\xx{10}$\,m $=5.6\xx{-7}$\,pc. In units of
$\rg(u_0)= m_p u_0/(e B_0)$, the units used for the $x$-coordinates in
Figs.~\ref{BL_noBL} and \ref{vary_Bz}, this corresponds to $|\Dfeb| =
1000$, $100$, and $10\,\rg(u_0)$, for $B_0 = 0.3$, $3$, and $30$~\muG,
respectively.
The top four panels showing $u(x)/u_0$ and energy flux have the same
format as the corresponding panels in Figure~\ref{BL_noBL}. As in
Figure~\ref{BL_noBL}, $\Qesc$ is significant and $\Rtot > 7$ in all
cases. The magnetic field panels differ from Figure~\ref{BL_noBL} in that
here we plot $B(x)/B_0$ and it's clear that the amplification of $B$ is
greatest for the lowest $B_0$, i.e., $B(x)/B_0$ increases with
increasing $\Malf$. For the examples shown here,
$B_2/B_0 \simeq 400$ for $B_0=0.3$~\muG,
$B_2/B_0 \simeq 150$ for $B_0=3$~\muG, and
$B_2/B_0 \simeq 30$ for $B_0=30$~\muG.
In the bottom panels, we show the pressure in magnetic turbulence,
$\Pwtot$, divided by the total far upstream momentum flux, $P_0$.  For
these examples, $\Pwtot/P_0 \lesssim 0.1$ and the magnetic pressure
stays well below equipartition with the gas pressure.

In Figure~\ref{vary_Bz_fp}a we show the distribution functions
corresponding to the shocks shown in Figure~\ref{vary_Bz}. As expected,
the shock with the highest $B_0=30$\,\muG\ yields the highest
$\Pmax$. This $\Pmax$, however, is only about a factor of 5 greater than
that for $B_0 = 0.3$\,\muG\ obtained in a shock system of the same
physical size. 
In Figure~\ref{vary_Bz_fp}b we show the distribution functions for three
cases where we have kept the upstream FEB boundary at the same number of
gyroradii, i.e., $\Dfeb=-100\rg(u_0)$. In this case, the situation with
$\Pmax$ is reversed with the $B_0=0.3$\,\muG\ shock obtaining the
highest $\Pmax$. This is a combination of the fact that the physical
size of the shock system is largest and that the field amplification is
greatest for $B_0=0.3$\,\muG. The shock structure results are not shown
but are similar to
those in Figure~\ref{vary_Bz}.

\subsection{Wave Amplification factor, $\fValf$}
All of the examples shown so far have used the minimum amplification
factor $\fValf=0$ (equation~\ref{VG}). We now investigate the effects of
varying $\fValf$ between 0 and 1 so that $V_G$ varies between $v_a(x)$
and $\Beff(x)/\sqrt{4 \pi \rho(x)}$. The other shock parameters are the
same as used for the dashed curves in Figure~\ref{BL_noBL}, i.e.,
$u_0=5000$\,\kmps, $B_0=30$\,\muG, and $\Dfeb=-1000\,\rg(u_0)$.

\begin{table}
\begin{center}
\caption{Effect of varying $\fValf$. The errors on $\Rtot$ and $\Rsub$
  are typically $\pm 10\%$.  
\vskip12pt
\label{table-1}}
\begin{tabular}{crrrrrrrrrrr}
\tableline
\tableline
$\fValf$
&$\Rtot$
&$\Rsub$
&$B_2/B_0$
&$p/(m_pc)_\mathrm{peak}^\mathrm{DS}$
\\
\tableline
0
&9
&3.5
&30
&$3.8\xx{-3}$
\\
0.1
&8
&3.7
&40
&$4.4\xx{-3}$
\\
0.5
&6
&4.1
&50
&$6.3\xx{-3}$
\\
1
&5
&4.3
&40
&$8.0\xx{-3}$
\\
\tableline
\end{tabular}
\end{center}
\label{numbers}
\end{table}

\begin{figure}        
\epsscale{.65} \plotone{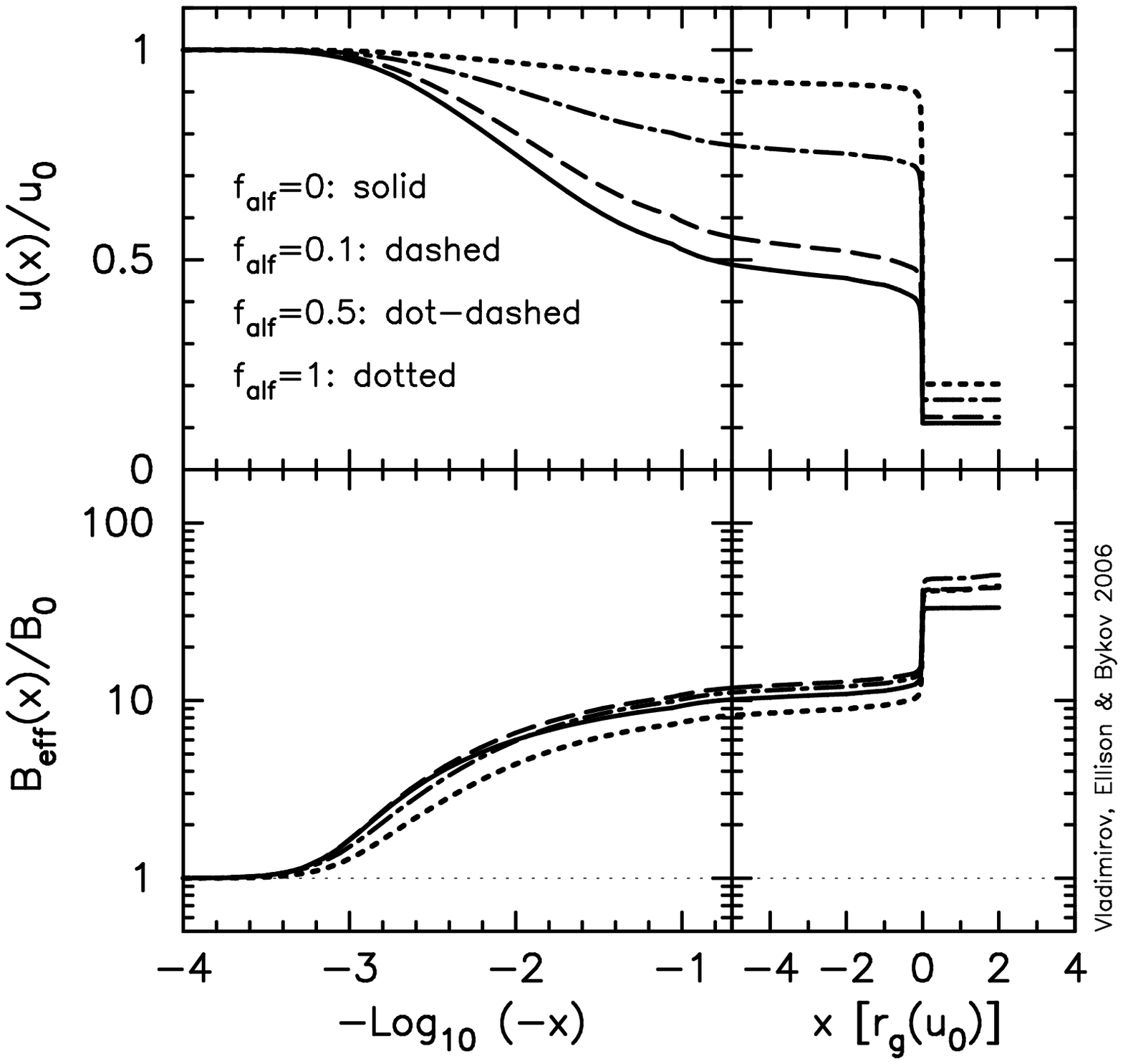}
\caption{Shocks with varying $\fValf$ as indicated. In all cases,
  $u_0=5000$\,\kmps, $B_0=30$\,\muG, and $\Dfeb=-1000\,\rg(u_0)$.
\label{amp_grid}}
\end{figure}

\begin{figure}        
\epsscale{.65} \plotone{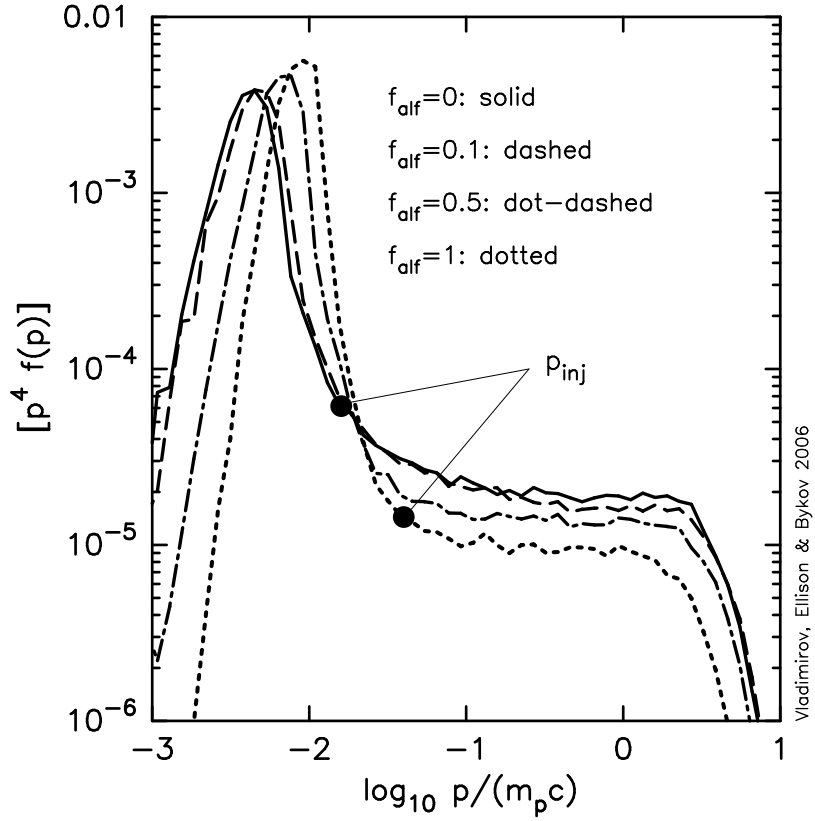}
\caption{Distribution functions for the shocks shown in
Figure~\ref{amp_grid} calculated downstream from the shock in the shock
reference frame. The solid dots give the approximate position for the
transition between ``thermal'' and superthermal particles for the two
extreme cases of $\fValf=0$ and $1$.
\label{amp_fp}}
\end{figure}

Figure~\ref{amp_grid} shows $u(x)/u_0$ and $\Beff(x)/B_0$ for $\fValf =
0$, $0.1$, $0.5$, and 1 as indicated.  The top panels show that
increasing the growth rate (increasing $\fValf$ and therefore $V_G$)
produces a large change in the shock structure and causes the overall
shock compression ratio, $\Rtot$, to decrease. 
The decrease in $\Rtot$
signifies a decrease in the acceleration efficiency 
and a decrease in the fraction of energy that escapes at the FEB, and
the subshock compression adjusts to ensure conservation of momentum and
energy. The values for $\Rtot$ and $\Rsub$ are given in Table~1 and it
is interesting to note that $\Rsub$ increases as $\Rtot$ decreases and
becomes greater than 4 for $\fValf \gtrsim 0.5$. 
In contrast to the strong modification of $u(x)$, there is little
difference in $\Beff(x)/B_0$ (bottom panels of Figure~\ref{amp_grid}) and
little change in $\Pmax$ (Figure~\ref{amp_fp}), between these examples.

The fact that increasing the wave growth rate decreases the acceleration
efficiency shows the \NL\ nature of the wave generation process and
comes about for two reasons.  The most important is that the magnetic
pressure term in equation~(\ref{pflux}), $\Pwtot$, becomes
significant compared to $\rho(x)u^2(x)$ when $\fValf \rightarrow 1$. 
The wave pressure causes the shock to be less compressive overall and
forces $\Rtot$ down.  Any change in $\Rtot$ changes the acceleration
efficiency and therefore changes $\dPCRxp$, the wave growth, and the
shock structure.
The second reason is that as the turbulence grows, $\Vwt$ grows and the
effective scattering center speed $u(x) + \Vwt$ drops, causing
the acceleration efficiency to drop. For the examples shown in
Figure~\ref{amp_grid}, the change produced by $\Vwt$ is modest compared to
the effect from $\Pwtot$. Taken together, at least for the parameters
used here, the increase in $\Pwtot$ and $\Vwt$ outweigh any increase in
efficiency the amplified field produces.

In Figure~\ref{amp_fp} we show the distribution functions for the four
examples of Figure~\ref{amp_grid} and note that the low momentum peaks
shift upward significantly with increasing $\fValf$. The approximate
momenta, measured in the downstream plasma frame, 
where the distributions peak, $p/(m_pc)_\mathrm{peak}^\mathrm{DS}$, are
listed in Table~1.
As we have emphasized, the injection efficiency, i.e., the fraction of
particles that enter the Fermi process, must adjust to conserve momentum
and energy and the low momentum peaks shift as a result of this.
The solid dots in Figure~\ref{amp_fp} roughly indicate
the injection point separating ``thermal''
and superthermal particles for the two extreme cases of $\fValf = 0$ and 1. 
The first thing to note is that this injection point is not well
defined, a consequence of the fact that the MC model
doesn't distinguish between ``thermal'' and ``nonthermal''
particles. Once the shock has become smooth, the injection process is
smooth and the superthermal population smoothly emerges from the
quasi-thermal population.\footnote{We note that the smooth emergence of
  a superthermal tail has been seen is spacecraft observations of the
  quasi-parallel Earth bow shock \citep[i.e.,][]{EMP90} and at
  interplanetary traveling shocks \citep[i.e.,][]{BOEF97}.} 
Nevertheless, the approximate momentum where the superthermal population
develops, $\pinj$, can be estimated and we mark this position with solid
dots for $\fValf=0$ and $1$.
What is illustrated by this is that the injection point shifts,
{\it relative to the post-shock distribution}, when $\fValf$ is varied.
This implies that, if injection is parameterized, the parameterization
must somehow be connected to modifications in the shock structure.

\section{SUMMARY AND COMPARISON WITH ALTERNATIVE MODELS}
We have generalized a steady-state \MC\ model of efficient diffusive
shock acceleration to include magnetic wave growth, allowing the wave
amplitude to become large compared to the ambient field, i.e., $\Delta
B/B_0 \gg 1$.  
The model uses a
phenomenological 
treatment of wave generation by applying the linear growth rate
formalism in the non-linear regime, but
couples the \NL\ shock structure, the injection rate of thermal
particles, the \MFA, and the determination of the maximum momentum
particles obtain from a physical constraint, i.e., the size of the shock
system, in an internally \SC\ manner.

Important limitations of our model are the simplified treatment of wave
growth and the neglect of wave damping. We also assume a Bohm-like
expression for the scattering mean free path [i.e., $\lambda(x,p)=
pc/(e\Beff)]$ rather than calculating $D(x,p)$ from more fundamental
relations which account for particle motions in strongly anisotropic
turbulence.
More physically realistic models for turbulence generation and damping
and particle scattering are certainly necessary and may well modify the
results we present here. However, it is not straightforward to include
these processes in \NL\ models and, to our knowledge, none have yet been
presented with as extensive \NL\ coupling as we calculate.
We believe the approximations we make are an important intermediate
step and that the results we present are indicative of what more
complete models will show. The fact that the MC technique can handle
anisotropic particle distributions
will be essential for 
including more precise plasma physics 
in future generalizations.
 
\subsection{Particle-In-Cell Simulations}
The difference between our \MC\ simulation and a
particle-in-cell (PIC) plasma simulation is that the pitch-angle
diffusion is treated phenomenologically in the MC with an assumed
scattering mean free path, rather than calculating particle trajectories
in the self-generated magnetic field obtained directly from Maxwell's
equations. This approximation is essential in order to make the MC
technique fast enough to model acceleration over a wide dynamic range.
We emphasize that while PIC simulations, in principle, can solve the
shock acceleration problem completely, their computational
requirements are extreme and they are not yet capable of modeling
acceleration over a dynamic range large enough to model cosmic sources
such as SNRs.
The most important constraint on PIC simulations is that
they must be done fully in
three dimensions.
As shown by \citet{JKG93} and \citet{JJB98}, PIC simulations with one or
more ignorable dimensions artificially confine particles to field lines
and eliminate cross-field diffusion, an essential ingredient in
diffusive shock acceleration, particularly in oblique shocks.
The 3-D requirement means that all three box dimensions
must be increased to accommodate high-energy particles with long
diffusion length scales
so that computing
requirements become insurmountable for a large enough dynamic range.
To model SNR shocks, they must be able to accelerate particles from
thermal to highly \rel\ energies and, if electrons are to be modeled,
they must simultaneously include electron and proton scales.\footnote{We
note that the computational requirements for the \rel\ shocks expected
in \gamray-bursts may actually be less stringent than those for the
\nonrel\ shocks in SNRs. In shocks with large Lorentz factors, particles
start off \rel\ (in the shock frame) and can gain a great deal of energy
in just a few shock interactions. It is also possible that
electron--positron plasmas dominate the \gamray-burst fireball so that
important results can be obtained without simultaneously covering
electron--proton scales.  In the \nonrel\ shocks present in SNRs,
however, both electrons and protons must be accelerated over a wide
dynamic range from eV to TeV energies by crossing the shock many times.}
While PIC simulations will be able to investigate important problems,
particularly those concerning injection, they will not be able to model
the shock acceleration of electrons and protons to the energies
necessary to produce broad-band radiation with parameters similar to
those of SNRs in the foreseeable future.  Until then, progress can be
made with approximate methods.

\subsection{Semi-Analytic Models with B-field Amplification}
Besides the limited results from PIC simulations, the only models
of magnetic field generation with $\Delta B \gg B_0$ in \NL\ DSA
that we are aware of are the \SA\ results of \citet{BL2001},
\citet{Bell2004,Bell2005}, \citet{PZ2003}, and \citet{AB2006}. As
we have mentioned, our work uses the same basic wave generation
formalism as \citet{BL2001}, with an extensive generalization to
include injection, \NL\ shock structure and particle
distributions, and $\Pmax$.

More recently, \citet{Bell2004,Bell2005} has attempted to improve
the plasma physics by calculating amplified fields from a
so-called ``directly driven'' mode of wave
instability.\footnote{\citet{LB2000} and \citet{Bell2004,Bell2005}
also performed PIC simulations coupled to a 3-D MHD model of the
background plasma. These results clearly showed that seed
$B$-fields can be amplified by orders of magnitude but they were
limited in dynamic range and did not \SCly\ model the particle
acceleration process.}
In this scheme, upstream energetic particles standing in the shock
frame produce a macroscopic current, and magnetic field, in the
upstream plasma frame.
This extended the \citet{BL2001} analysis to wave modes other than
resonant Alfv\'enic ones and emphasized that turbulence is likely
the result of strongly driven, non-resonant modes at shorter
wavelengths rather than \alf\ waves.
A recent application of this idea to \rel\ shocks expected in GRBs is
given \citet{MN2006}.
While this is a promising idea for a non-resonant mechanism for
wave amplification in strong shocks, careful consideration of the
plasma return current must be made (e.g., F.C. Jones, private
communication).

Another non-resonant mechanism for long-wavelength magnetic field
fluctuation amplification in the vicinity of shock waves was studied by
\citet{BT2005}.  They demonstrated that a small fraction of neutral
atoms can reduce the plasma transverse conductivity and result in a
cosmic-ray current instability.  This instability was shown to produce
long-wavelength magnetic fluctuations.

An important attempt at describing turbulence with $\Delta B/B \gg
1$ was made by \citet{PZ2003,PZ2005}. These authors assumed a
Kolmogorov-type \NL\ cascade and damping of self-generated
turbulence by ion-neutral collisions. Most importantly, they
obtained estimates for $\Pmax$ in a time-dependent analytic
calculation. Nonlinear particle acceleration was assumed, but the
analytic method required a number of approximations including the
spatial distribution of energetic particles in the shock precursor
and the spectral form of the energetic particles. Nevertheless,
this work showed that \NL\ particle acceleration, combined with
\BFA, may strongly influence $\Pmax$, one of the most important
parameters in diffusive shock acceleration.

In our estimation, the most highly developed model of \NL\ DSA with
\MFA\ is that of \citet{AB2006}; work based on a series of papers by
Blasi and co-workers
\citep[i.e.,][]{Blasi2002,Blasi2004,BGV2005,AB2005}.\footnote{We note
that \citet{AB2005,AB2006} use a different technique from the previous
\Blasi\ work that allows an exact solution for an arbitrary choice of
both the spatial and momentum dependent diffusion coefficient.}
We now give a detailed comparison between that model and ours.
The underlying assumptions for DSA are the same in both models, i.e.,
particles are driven to isotropy by interactions with magnetic
turbulence in the background plasma. Particles gain energy when they
diffuse in the converging flow and
neither
model includes second-order Fermi acceleration. Both models are also for
plane-parallel, steady-state shocks.

The most important difference revolves around the method
of solution. We use a \MC\ technique while Blasi et al. solve a
transport or diffusion-convection equation.
In the \Blasi\ work, macroscopic quantities (pressure, energy flux,
etc.) are derived as moments of the particle distribution function
$f(x,p)$, which is assumed to be isotropic in the shock reference
frame. In contrast, the \MC\ simulation traces the stochastic motion of
individual particles as they pitch-angle scatter off the background
turbulence and calculates $f(x,p)$ and its moments directly from the
particle trajectories without making any assumptions about the isotropy
of $f(x,p)$. In effect, the MC simulation solves the more fundamental
Boltzmann equation \citep[e.g.,][]{EE84}.
Of course, the semi-analytic technique is much faster
computationally than the \MC\ technique and this will be important for
building models  in many applications.

Injection can be treated \SCly\ in the MC simulation because, once the
pitch-angle scattering assumptions are made, they are applied equally to
all particles and the number and energy of injected particles is fully
determined.  No distinction is made between thermal and superthermal
particles and the viscous subshock is assumed to be transparent so there
is a nonzero probability for any downstream particle with $v>u_2$ to be
injected.
The diffusion approximation, on the other hand, with its assumption of
isotropy forces additional assumptions if injection is to be modeled.
\Blasi\ treat the subshock as having a finite thickness comparable to a
thermal particle's gyroradius and the injection rate is parameterized
such that only those particles get injected, which have a gyroradius
large enough to span the subshock.
While this adds an additional parameter independent of the diffusion
properties, it has been shown that parameters can easily be determined
so that the \MC\ and semi-analytic models give similar results
\citep[see][]{EBG2005}. 

Another important difference is that
\citet{AB2006} have included a
phenomenological description of turbulent $B$-field heating, similar in
implementation to that used in \citet{BE99}, in addition to adiabatic
heating. We only
include adiabatic heating.
The amplified turbulence may be dissipated through collisional and/or
collisionless mechanisms and these include:
\newlistroman
\listromanDE linear and \NL\ Landau damping 
\citep[e.g.,][]{AEA75,AB86,Kulsr78,VBT93,Z2000},
\listromanDE particle trapping \citep[e.g.,][]{Medvedev99},
and
\listromanDE ion-neutral wave damping \citep[e.g.,][]{DDK96,BT2005}.  
It's important to note that even though the damping rates tend to be
smaller than the wave growth rates for long-wavelength fluctuations in
turbulent plasmas with a substantial nonthermal component
\citep[e.g.,][]{BT2005}, the generation of magnetic energy density can
be approximately balanced by the convection of that turbulence through
the shock (as was assumed here) and MHD cascade processes. The heating
of the precursor plasma by dissipation of small scale fluctuations
modifies the subshock Mach number \citep[e.g.,][]{EBB2000} and this in
turn modifies injection.
The
overall acceleration efficiency and, of particular importance for X-ray
observations, the temperature of the shocked plasma
\citep[e.g.,][]{DEB2000,HRD2000,EDB2004} will depend on wave dissipation.
We plan to include physically realistic models of wave damping in the MC
simulation in future work.

The two most important elements in these \NL\ models is the turbulence
generation and the diffusion coefficient that is derived from it.  The
assumptions for wave generation by the streaming instability are
essentially identical in the \Blasi\ model and ours.\footnote{There is a
minor difference in that \citet{AB2006} assume that all waves generated
by the streaming instability move upstream whereas we consider the
interaction between waves moving upstream and downstream.}
The most important difference in the models lies in the calculation of
the mean free path and diffusion coefficient. Here, we assume 
Bohm-like diffusion with $\Beff$ (equation~\ref{mfp}), while
\citet{AB2006} use a mean free path due to resonant scattering that, in
our notation, would be
\begin{equation}
\label{mfp_ab06}
\Lres(x,p) = \frac{cp}{e B_0} \frac{1}{2 \pi^2 \Kres}
\frac{B_0^2}{\left [ U_-(x,\Kres) + U_+(x,\Kres) \right ]} 
\ ,
\end{equation}
where 
\begin{equation}
\Kres = \frac{1}{\rgZ} = \frac{e B_0}{cp}
\end{equation}
is the resonant wavenumber.  Both of these assumptions are
approximations. Our Bohm-like expression is just a phenomenological
recipe for particle diffusion in strong turbulence, while $\Lres$ is
only formally valid for weak turbulence but is applied to diffusion in
strong, anisotropic turbulence.  An important
step in advancing the state-of-the-art of \NL\ shock acceleration must
center on improving the connection from wave generation to diffusion.

\begin{figure}        
\epsscale{.65} \plotone{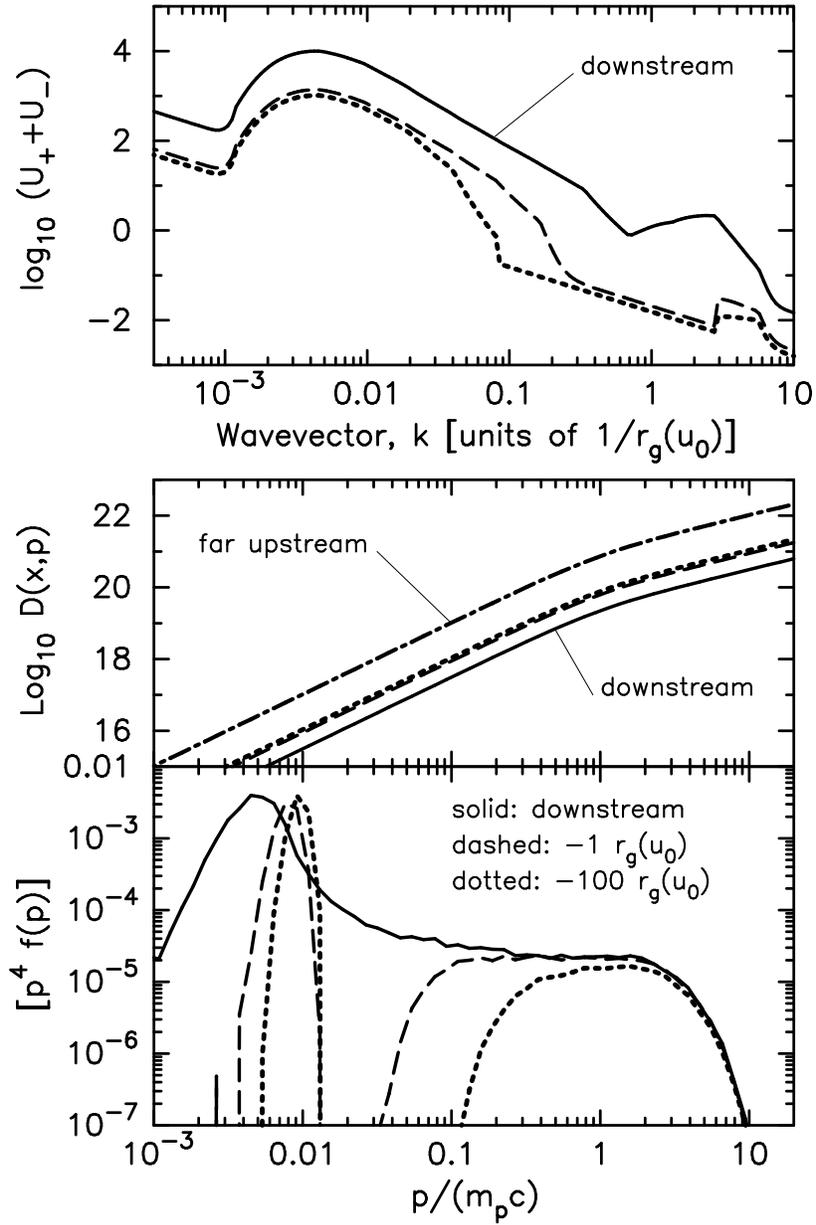}
\caption{The top panel shows $U_+(x,k) + U_-(x,k)$ vs. $k$ at three
positions relative to the subshock. Here and in the other two panels,
the solid curve is calculated downstream from the shock, the dashed
curve is calculated at $x = - \rg(u_0)$ upstream from the subshock, and
the dotted curve is calculated at $x = -100\rg(u_0)$ upstream from the
subshock. The middle panel shows the diffusion coefficient with an
additional dash-dotted curve showing the far upstream value. The bottom
panel shows the distribution functions, multiplied by $[p/(m_pc)]^4$, at
the various positions. These distributions are calculated in the shock
frame.  
\label{U_D_fp}}
\end{figure}

That important aspects of DSA are sensitive to the assumptions made for
diffusion can be seen by comparing the self-generated $\Diff$ from these
two models.
Our use of equation~(\ref{mfp}) forces Bohm--like diffusion with
$\Diff(x,p) \propto v p$ and we show this dependence in
Figure~\ref{U_D_fp}. Nevertheless, we have argued in
Section~\ref{Sect_scat} that equation~(\ref{mfp}) overestimates the
scattering strength for the highest momentum particles because most of
the harmonics appear as short-scale fluctuations to these particles.
This would suggest that the actual $p$ dependence of $\Diff$ increases
with $p$, likely as $\Diff \propto p^2$ at the highest energies.
In \citet{AB2006}, on the other hand, the momentum dependence of $\Diff$
is shown to weaken at high $p$ 
to the point where the scattering becomes strong enough that
$\Diff$ becomes
independent of $p$ in high sonic Mach number shocks.

It is worth noting that the \citet{AB2006} result is indicating that the
magnetic turbulence spectrum has a slope approaching $\propto
k^{-2}$. For this turbulence spectral slope, the {\it quasi-linear
model} predicts energy independent diffusion (in the weak fluctuation
limit of course). Interestingly enough, the magnetic fluctuation
spectrum we obtain with Bohm diffusion has, over a wide wavenumber
range, a spectral slope roughly corresponding to $\propto k^{-2}$ (see
Figure~\ref{U_D_fp}).
So, if one formally calculated the {\it quasi-linear} diffusion from
equation~(\ref{mfp_ab06}) with our fluctuation spectrum, the resulting
diffusion coefficient would have only a weak energy dependence. This
could mean that the magnetic fluctuation spectral shape is not very
sensitive to the choice of diffusion model in the nonlinear
calculations. Nevertheless, in this preliminary work, we prefer to use
the Bohm diffusion model in the strong fluctuation limit.
Since the behavior of $\Diff$ at high $p$ determines $\Pmax$, as well as
influencing all other aspects of the shock acceleration process, 
it's clear that much work remains to be done on this difficult problem.

\subsection{Other Models of Non-linear Shock Acceleration 
Without B-field Amplification}
There are a number of \NL\ models of DSA besides those discussed
above. However, none of these models include $B$-field amplification.

In a series of papers, Berezhko and co-workers have applied a model of
\NL\ DSA to broad-band observations of several young SNRs
\citep[e.g.,][]{BEK96a,BKV2002}.  They used a time-dependent solution of
the cosmic-ray transport equation coupled to the gas dynamic equations
in spherical symmetry and calculated the superthermal proton
distribution, from the forward shock, at all positions in the
remnant. From this they can determine the overall contribution a single
SNR makes to the galactic \CR\ proton flux.  Furthermore, since they
included the acceleration of superthermal electrons, they calculated the
photon emission from \syn\ and IC processes, as well as from
proton-proton interactions and \pion. This allowed them to fit the
broad-band photon observations and constrain the model parameters. Of
particular importance is their emphasis that magnetic fields much
greater than typical ISM values are required to match broad-band photon
observations. Equally important is their prediction, based on their
fits, that TeV photon emission from several SNRs is most likely the
result of \pion\ from protons rather than IC from electrons
\citep[e.g.,][]{BKV2003}.

This model has been extremely successful and has added to our
understanding of SNRs, but it does make important
approximations and simplifications. 
The model assumes Bohm diffusion, parameterizes the number of injected
particles, and does not include wave amplification.\footnote{That
$B$-field amplification is {\it not} included in the \Berezhko\ model
can easily be missed, particularly since the titles of some of these
papers, e.g., ``Confirmation of strong magnetic field amplification and
nuclear cosmic ray acceleration in SN 1006'' \citep[][]{BKV2003}, might
suggest that the model does contain $B$-field amplification. The model
of \Berezhko\ is a parallel shock model where the magnetic field is {\it
not} explicitly included in the convection-diffusion equations. There is
no $B$-field amplification and the large downstream fields that
\Berezhko\ infer from matching the observations are obtained by an ad
hoc compression of the upstream field \citep[see, for example, the
discussion before equation (8) in][]{VBKR2002}. To obtain 300\,\muG\
downstream, for example, they must start with an unshocked field of
$\sim 50$\,\muG\ which is then compressed by the shock with a ratio
typically $\Rtot \sim 6$
for their models.}

Other \NL\ shock models include those of Kang and Jones and co-workers
\citep[e.g.,][]{KJG2002,KJ2006}, Malkov and co-workers
\citep[e.g.,][]{Malkov97,MD2001,MDJ2002,MD2006}, and the cosmic
ray--hydrodynamical (CR-hydro) model of Ellison and co-workers
\citep[e.g.,][]{EDB2004}.
These models
all involve different computational techniques, and all have their
particular strengths and weaknesses. They have also been shown to produce
similar \NL\ effects to those of \Berezhko, and have been shown to
be in quantitative agreement with our \MC\ model (before the addition of
$B$-amplification).
Since none of these models yet include \BFA\ we do not discuss them
further.

\section{CONCLUSIONS}
We have introduced a model of diffusive shock acceleration which couples
thermal particle injection, \NL\ shock structure, \MFA, and the \SC\
determination of the maximum particle momentum. This is a first step
toward a more complete solution and, in this preliminary work, we make
a number of approximations dealing mainly with the plasma physics of
wave growth. 
Keeping in mind that our results are subject to the validity of our
approximations, we reach a number of interesting conclusions.

First, our calculations find that efficient shock acceleration can
amplify ambient magnetic fields by large factors and are generally
consistent with the large fields believed to exist at blast waves in
young SNRs, although we have not attempted a detailed fit to SNR
observations in this paper.
While the numerical values we obtain depend on the particular parameters
for our examples, and we will investigate in detail how the
amplification depends on sonic Mach number, age, and size of the shock
system, etc., in future work, amplification factors of several 100 are
clearly possible.

\begin{figure}        
\epsscale{.65} \plotone{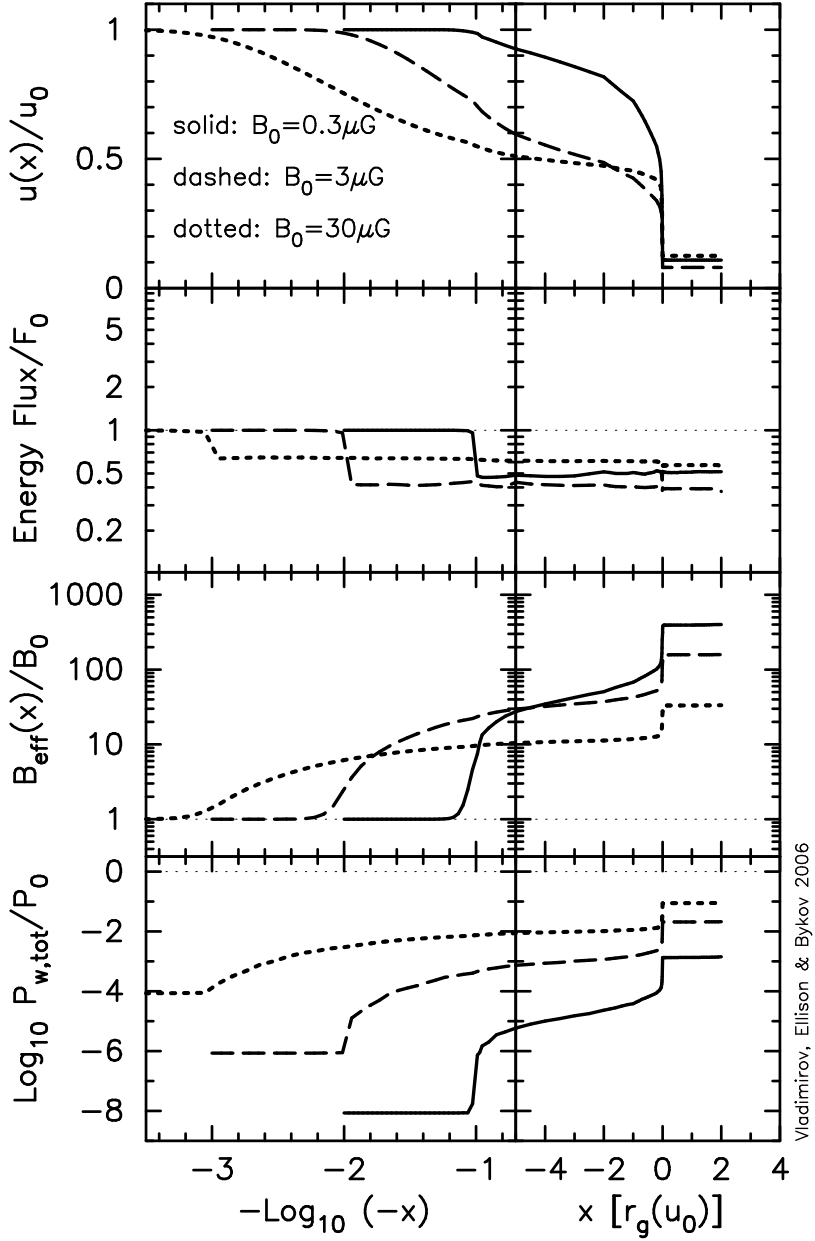}
\caption{Comparison of shocks with different far upstream fields
  $B_0$. In all panels, the solid curves are for $B_0=0.3$\,\muG, the
  dashed curves are for $B_0=3$\,\muG, and the dotted curves are for
  $B_0=30$\,\muG. The FEB is placed at the same physical distance in all
  cases with $\Dfeb=-1.7\xx{10}$\,m. The horizontal axis is in units of
  $\rg(u_0)\equiv m_p u_0/(e B_0)$, and is split at $x=-5\rg(u_0)$ between
  a linear and logarithmic scale. Note that $\Beff$ increases most
  strongly for $B_0=0.3$\,\muG, but that the pressure in magnetic
  turbulence never gets above $\sim 10$\% of the total pressure. The
  overall compression ratios are: 
$\Rtot \simeq 9$ for $B_0=0.3$\,\muG,    
$\Rtot \simeq 12$ for $B_0=3$\,\muG,    
$\Rtot \simeq 8$ for $B_0=30$\,\muG, values consistent, within
  statistical errors, with $\Qesc$, as indicated in the energy flux
  panels.
\label{vary_Bz}}
\end{figure}

More specifically, we find that the amplification, in terms of the
downstream to far upstream field ratio $B_2/B_0$ is a strong function of
\alf\ Mach number, with weak ambient fields being amplified more than
strong ones. For the range of examples shown in Figure~\ref{vary_Bz},
$B_2/B_0 \sim 30$ for $\Malf \sim 80$ and $B_2/B_0 \sim 400$ for $\Malf
\sim 8000$. 
Qualitatively, a strong correlation between amplification and $\Malf$
should not depend strongly on our approximations and may have important
consequences.
In young SNRs, the expansion of the ejecta material will drastically
reduce any original, pre-SN circumstellar magnetic field. In fact, for
any conceivable progenitor, the magnetic field inside of the reverse
shock will drop to values too low to support the acceleration of
electrons to radio emitting energies only a few years after the
explosion \citep[e.g.,][]{EDB2005}. Evidence for radio emission at
reverse shocks in SNRs has been reported \citep[see,][for
example]{GotthelfEtal2001} and the strong amplification of low fields we
see here, may make it possible for reverse shocks in young SNRs to
accelerate electrons to \rel\ energies and produce radio \syn\ emission.
If similar effects occur in \rel\ shocks, these large amplification
factors will be critical for the internal shocks presumed to exist in
\gamray\ bursts (GRBs). Even if large \BFA\ is confined to \nonrel\
shocks, which tend to be more efficient accelerators than \rel\ ones,
amplification will be important for understanding GRB afterglows since
the expanding fireball will slow as it moves through the interstellar
medium and will always go through \transrel\ and \nonrel\ phases.

\begin{figure}        
\epsscale{.65} \plotone{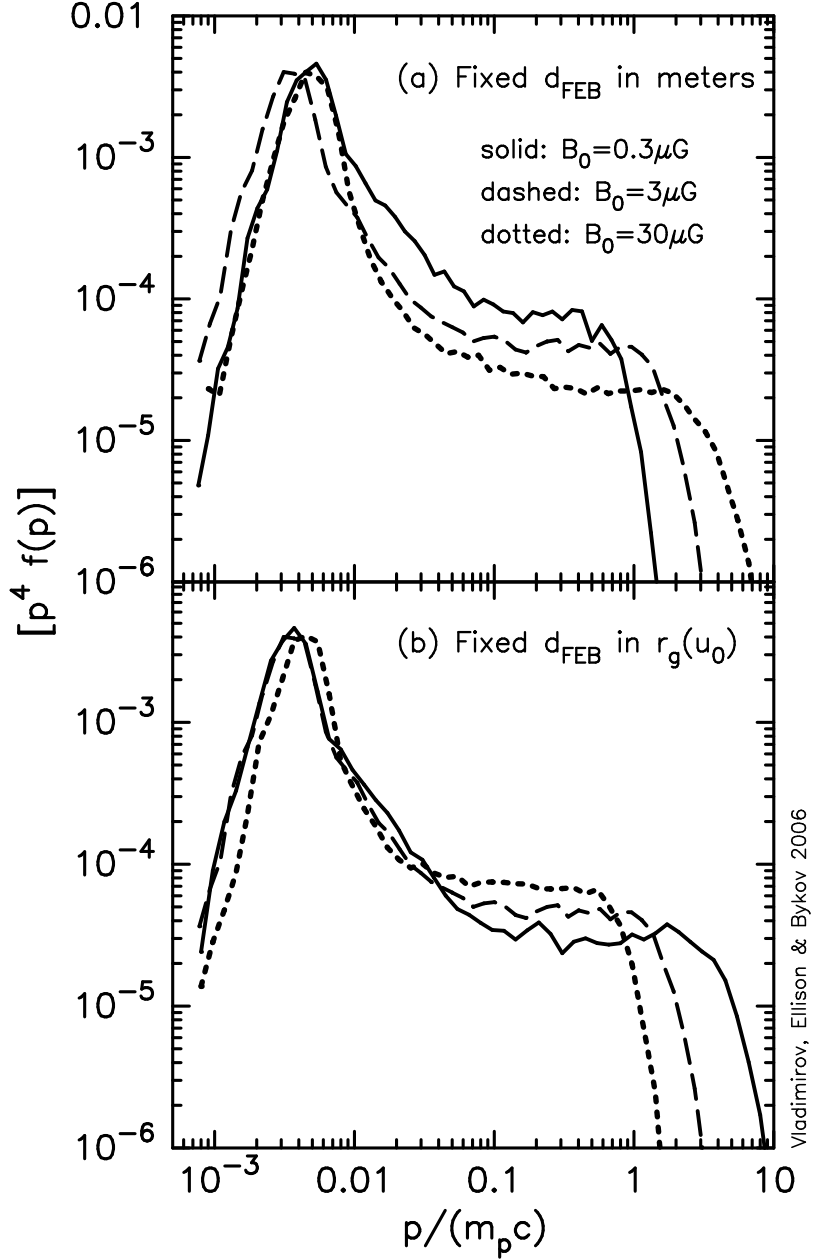}
\caption{The top panel shows the distribution functions obtained in the
  shock frame in the downstream regions of the shocks shown in
  Figure~\ref{vary_Bz}.  These shocks have the FEB at the same physical
  distance from the subshock and the shock with $B_0=30$\,\muG\ produces
  the greatest $\Pmax$. In the bottom panel we show a similar set of
  curves only here the FEB was set at a fixed $\Dfeb = -100\rg(u_0)$
  upstream. In this case, the shock with $B_0=0.3$\,\muG\ produces the
  greatest $\Pmax$, a result of the larger shock size and greater
  amplification factor the high $\Malf$ shock receives.
  \label{vary_Bz_fp}}
\end{figure}

As expected, amplifying the magnetic field leads to a greater maximum
particle momentum, $\Pmax$, a given shock can produce. Quantifying
$\Pmax$ is one of the outstanding problems in shock physics because of
the difficulty in obtaining parameters for typical SNRs that allow the
production of cosmic rays to energies at and above the CR knee near
$10^{15}$\,eV. 
Assuming that acceleration is truncated by the size of the shock system,
we determine $\Pmax$ from a physical constraint: the relevant parameter
is the distance to the free escape boundary in diffusion lengths. This
means that the limit on acceleration feeds back on the shock structure
and also mimics, in terms of the spectral shape, what happens in actual
shocks where $f(x,p)$ must turn over smoothly at the highest energies
(as in Figure~\ref{vary_Bz_fp}). The spectral shape will be particularly
important if the model is applied to the knee of the cosmic-ray spectrum
or to nonthermal X-ray emission in SNRs.

\begin{figure}        
\epsscale{.65} \plotone{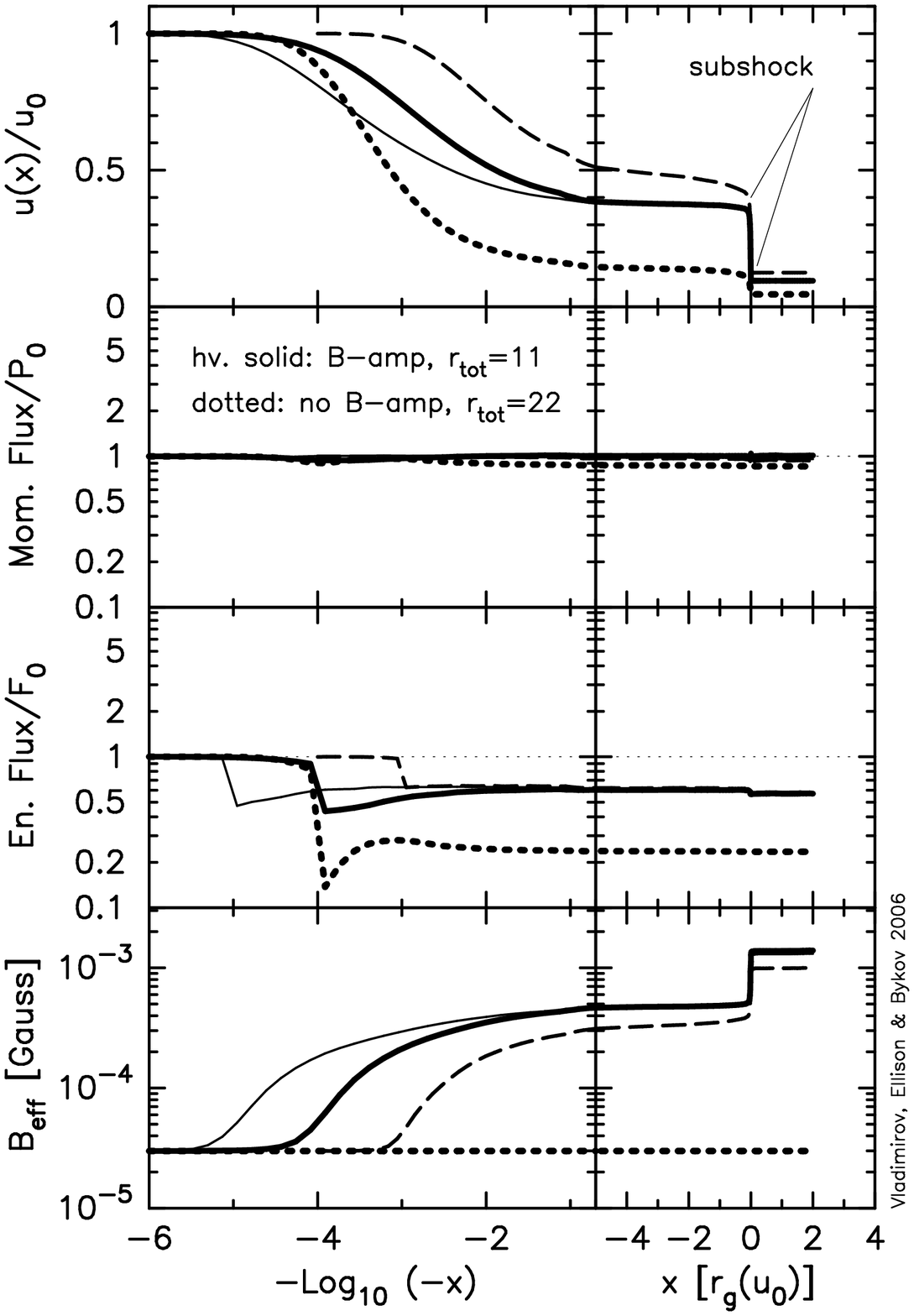}
\caption{Shock structure including momentum and energy fluxes (in units
  of far upstream values) and the effective magnetic field vs. $x$. Note
  that the horizontal scale has units of $\rg(u_0)=m_pu_0/(e B_0)$ and
  is divided at $x=-5\rg(u_0)$ between a linear and log scale.  In all
  panels, the heavy dotted curves show results without amplification and
  all other curves are with amplification.  The curves showing the
  energy flux drop sharply at the upstream FEB, which is at $-10^4
  \rg(u_0)$ for the heavy solid and dotted curves, at $-1000\rg(u_0)$
  for the dashed curves, and at $-10^5\rg(u_0)$ for the light solid
  curves, as particles freely leave the system. When this escaping flux
  is included, energy and momentum are conserved to within $\pm 10\%$.
\label{BL_noBL}}
\end{figure}

\begin{figure}        
\epsscale{.65} \plotone{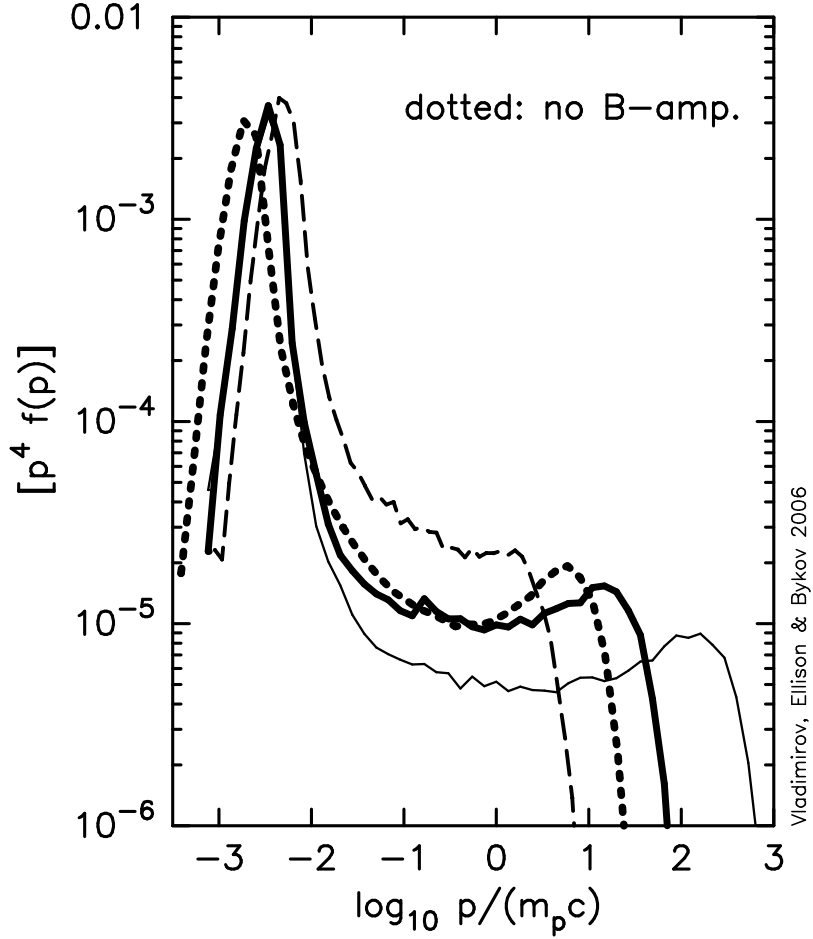}
\caption{Phase space distributions for the shocks shown in
  Figure~\ref{BL_noBL}. These spectra are multiplied by $[p/(m_pc)]^4$ and
  are calculated downstream from the shock in the shock rest frame. As
  in Figure~\ref{BL_noBL}, the heavy solid and dotted curves have
  $\Dfeb=-10^4\,\rg(u_0)$, the dashed curve has $\Dfeb=-1000\,\rg(u_0)$,
  the light solid curve has $\Dfeb=-10^5\,\rg(u_0)$.
\label{BL_noBL_fp}}
\end{figure}

Our results show that $\Pmax$ does increase when field amplification is
included, but the increase is considerably less than the amplification
factor at the shock $B_2/B_0$ (compare the heavy dotted and heavy solid
curves in Figure~\ref{BL_noBL_fp}).
The main reason for this is that high momentum particles have long
diffusion lengths and the precursor magnetic field well upstream from
the subshock strongly influences $\Pmax$. We calculate the spatial
structure of the amplified magnetic field (Figs.~\ref{BL_noBL} and
\ref{vary_Bz}) and, as expected, field amplification is greatest near
the subshock and $\Beff$ merges into the ambient field far upstream.
The diffusion length that determines $\Pmax$ (i.e.,
$[\DiffPmax/u(x)]_\mathrm{ave}$) is some weighted average over the
varying $u(x)$ and $\Beff(x)$ and is considerably greater than that
estimated from $B_2$ alone. If the shock size, in our case $\Dfeb$,
limits acceleration, $\Pmax$ will be considerably less than crude
estimates using a spatially independent $B_2$.

On the other hand, particles spend a large fraction of their time
downstream from the shock where the field is high and collision times
are short. If shock age limits acceleration rather than size, we expect
the increase in $\Pmax$ from the amplified field to be closer to the
amplification factor, $B_2/B_0$.
Thus, the spatial structure of the precursor field and $u(x)$, in
addition to the overall amplification of $B$, will determine the
relative time spent upstream versus downstream and will determine
$\Pmax$ for a given shock system. This will be even more critical for
electrons than for ions since electrons experience \syn\ and \IC\ losses
which will mainly occur downstream. 
Again, the qualitative nature of the above conclusions should not depend
on the particular parameters and approximations we make here. We leave
more detailed quantitative work for future study.

\begin{figure}        
\epsscale{.65} \plotone{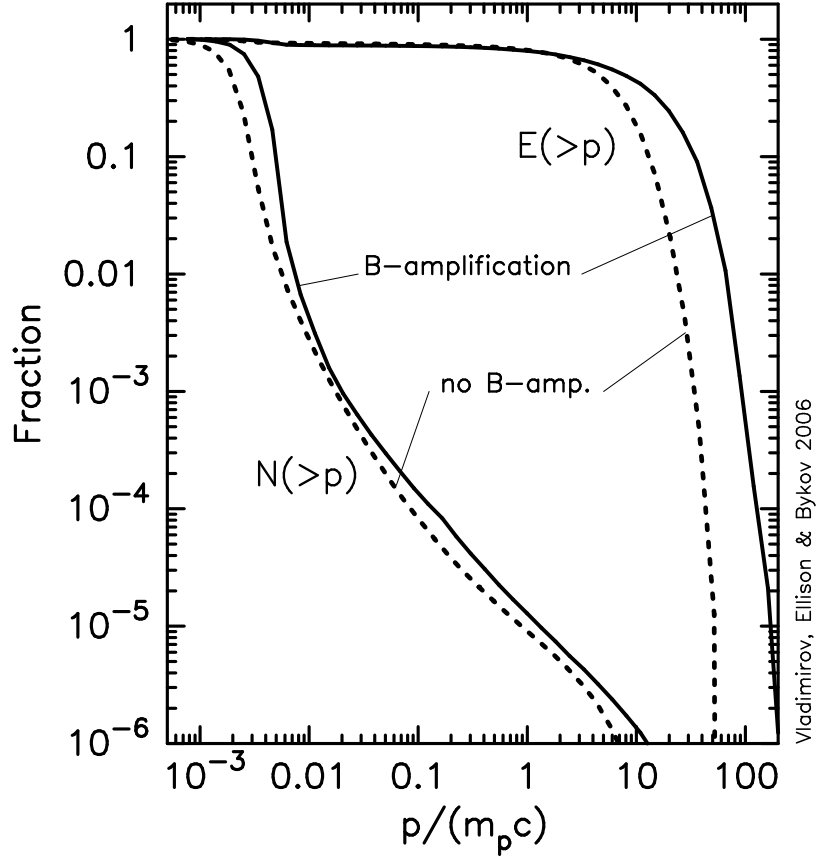}
\caption{The left-hand curves show the fraction of particles with
  momentum greater than $p$, while the right-hand curves show the
  fraction of energy density in particles with momentum greater than
  $p$, for the shocks shown in Figs.~\ref{BL_noBL} and \ref{BL_noBL_fp}
  with heavy solid and dotted curves. These curves are calculated from
  the distributions shown in Figure~\ref{BL_noBL_fp} and do not include
  the particles that escaped at $\Dfeb$. While the number fraction of
  escaping particles is small, the energy fraction is significant, as
  indicated in Figure~\ref{BL_noBL}. The solid curves show results with
  field amplification and the dotted curves show results without field
  amplification, both with $\Dfeb=-10^4\,\rg(u_0)$.
\label{inj_eff}}
\end{figure}

Finally, it is well known that DSA is inherently efficient. Field
amplification reduces the fraction of shock ram kinetic energy that is
placed in \rel\ particles but, at least for the limited examples we show
here, the overall acceleration process remains extremely efficient. Even
with large increases in $\Beff(x)$, well over 50\% of the shock energy can go
into \rel\ particles (Figure~\ref{inj_eff}).  As in all \SC\ calculations,
the injection efficiency must adjust to conserve momentum and energy. In
comparing shocks with and without field amplification, we find that
field amplification lowers $\Rtot$ and, therefore, individual energetic
particles are, on average, accelerated less efficiently. In order to
conserve momentum and energy, this means that more thermal particles
must be injected when amplification occurs. The shock accomplishes this
by establishing a strong subshock which not only injects a larger
fraction of particles, but also more strongly heats the downstream
plasma. This establishes a \NL\ connection between the field
amplification, the production of cosmic rays, and the X-ray emission
from the shocked heated plasma.

\acknowledgments
The authors are grateful to P. Blasi  and F.C. Jones for useful
discussions. DCE acknowledges support from a NASA ATP grant
(ATP02-0042-0006) and a NASA LTSA grant (NNH04Zss001N-LTSA), and AB was
supported, in part, by RBRF 06-02-16844.


\end{document}